\documentclass[pra,twocolumn,amsmath,amssymb,superscriptaddress]{revtex4-1}

\usepackage{graphicx}   
\usepackage{color}      
\usepackage{float}
\usepackage{braket}
\usepackage{bbm}
\usepackage{bbold}

\usepackage[caption=false]{subfig}
\usepackage{tkz-berge}
\usepackage{tikz}
\usetikzlibrary{calc,arrows}

\newcommand{\nn}{\nonumber}

\begin{document}

\title{Floquet engineering in superconducting circuits: from arbitrary spin-spin interactions to the Kitaev honeycomb model}

\author{Mahdi Sameti}
\affiliation{Institute of Photonics and Quantum Sciences, Heriot-Watt University Edinburgh EH14 4AS, United Kingdom}
\author{Michael J. Hartmann}
\affiliation{Institute of Photonics and Quantum Sciences, Heriot-Watt University Edinburgh EH14 4AS, United Kingdom}
\date{\today}

\begin{abstract}
We derive a theory for the generation of arbitrary spin-spin interactions in superconducting circuits via periodic time modulation of the individual qubits or the qubit-qubit interactions. The modulation frequencies in our approach are in the microwave or radio frequency regime so that the required fields can be generated with standard generators. Among others, our approach is suitable for generating spin lattices that exhibit quantum spin liquid behavior such as Kitaev's honeycomb model.
\end{abstract}


\maketitle

\section{Introduction}
Floquet theory provides a convinient framework to study periodically driven quantum systems. Manipulating slow dynamics of physical systems by applying periodic modulations, denoted as {\it Floquey engineering} \cite{Holthaus16,Oka18}, has turned into an active and rich area of research in recent years. This is motivated by the novel phenomena that can arise in these systems but are absent in their static counterparts. Following Floquet theory, the dynamics of periodically driven systems can, on longer time scales, typically be described in terms of an effective time-independent Hamiltonian. The idea of Floquet engineering is to shape the properties of this effective Hamiltonian by designing appropriate driving schemes.

Depending on the frequency of the external time modulation, one can distinguish between two different regimes of Floquet engineering. In the {\it off-resonance} or {\it high-frequency} regime the frequency of the modulation is much larger than the intrinsic frequencies of the system and the description of this regime is thus based on high-frequency or Magnus expansions \cite{Eckart15,Goldman14,Bukov15}. In contrast, in the {\it resonant modulation} regime, the modulation frequency is similar to the intrinsic frequencies of the system and the theory for high modulation frequencies is not directly applicable. For explicit resonances \cite{Goldman15}, one can transform the system to the rotating frame of the resonant interaction to obtain an effective high-frequency regime. Yet, the applicability of such an approach is not clear where {\it implicit} resonances happen in higher order of perturbation \cite{Sameti16}. In this work, we develop a Floquet engineering approach in the {\it resonant modulation} regime and apply it to coupled superconducting qubits.

In recent years, superconducting circuits have made tremendous advances in implementing engineered quantum dynamics in both, quantum information processing \cite{Barends14,Kelly15,Riste14,Corcoles15,Barends16,Ofek16} and quantum simulation \cite{Chen14,Raftery14,Barends15,Salathe15,Eichler15,Roushan16}. This technology has thus reached a stage of development that offers intriguing avenues for emulating and exploring quantum many-body physics \cite{Eichler15,Houck12,Fitzpatrick,Roushan17}. In superconducting circuits, Floquet theory has so far been applied to studty multi-photon transitions in a single Flux qubit \cite{Son09} and for quantum simulation of 1D spin chains via high-frequency modulation \cite{Kryiienko17}. Moreover, Floquet states of a superconducting qubit were experimentally observed \cite{Chunqing15}. 

Here we apply resonant Floquet engineering to superconducting circuits and present an approach for the analog quantum simulation of any spin-spin interactions of the form
\begin{equation} \label{eq:eff-ints}
	J e^{i\theta}\sigma_1^{\alpha}\sigma_2^{\beta} + \textrm{H.c.},
\end{equation} 
where $\alpha, \beta \in \{ x,y,z,+,-\}$ and $\sigma_j^\alpha$ are Pauli operators, $J$ the amplitude and $\theta$ the phase of the coupling constant. In particular, we extend Floquet engineering to the case of quasi-periodic (bimodal) driving. 

The strength of couplings between superconducting qubits is typically weak compared to the transition frequencies of the qubits. This leads to a simple hopping interaction that conserves the number of excitations and only becomes efficient if the transition frequencies of the qubits are tuned close to each other. Interactions that do not conserve the number of excitations are accessible in the ultrastrong coupling regime \cite{Wendin17}, but such coupling strengths lead to long-range interactions in many-qubit systems and the validity of a local mode description for large lattices is not clear. 

Arbitrary spin-spin interactions can also be engineered via digital quantum simulation where the target evolution is implemented as a Trotter sequence of elementary interaction steps. Yet, this approach requires an intricate gate sequence with an execution time that can exceed the coherence time of the qubits, particularly for multi-qubit systems. Employing shorter pulses does not solve this problem as such pulses are broader in frequency and may thus lead to unwanted resonances. Moreover, digital quantum simulation suffers from the error of the Trotter decomposition which hampers scalability to larger systems. 

Our single frequency or bimodal modulation approach, in contrast, realizes arbitrary spin-spin interaction in an analog manner in the rotating frame of the qubits, requiring only moderate interaction strength among the qubits. The engineered spin-spin interactions can be used as building blocks of analog quantum simulators with superconducting circuits which circumvent problems encountered in digital quantum simulation.

Our approach considers both, modulations applied to the individual qubits and to the coupling between them can be described. In the scheme with modulated couplings, one can obtain stronger interaction strength at the cost of a slightly higher complexity, as it requires additional control for the dynamical coupler compared to the driven qubit setup.

Our Floquet engineering is based on the seminal work by Shirley \cite{Shirley65}. This powerful method is particulary versatile since it can be easily generalized to multi-mode Floquet systems, where the external drive includes several frequencies \cite{Ho83, Leskes10}. We use a generalization of Salwen nearly degenerate perturbation theory \cite{Salwen55} to derive an effective Hamiltonian from the infinite-dimensional Floquet matrix. Using this method, we are able to analytically extract the parameters of the effective Hamiltonian to arbitrary orders of perturbation beyond rotating wave approximation.

The remainder of this paper is organized as follows. In section \ref{sec:theoretical-toolbox} we give an overview of the theoretical tools we employ, including single and multimode Floquet theory, adiabatic elimination and Salwen perturbation theory. The reader who is only interested in the results may directly go to sections \ref{sec:driven_qubit_schemes} and \ref{sec:driven_coupling} where we describe the spin-spin interactions that can be generated by driving the involved qubits or the coupling circuits. In section \ref{sec:Honeycomb} we then show how the generated interactions can be put together for realizing the honeycomb model by Kitaev  and conclude in section \ref{sec:conclusions}.

\section{Theoretical Toolbox}
\label{sec:theoretical-toolbox}
In this section, we introduce the theoretical toolbox that we later apply to various superconducting circuits.  We use Shirley's Floquet formalism for a single mode driven system \cite{Shirley65} and generailze it to many modes. This formalism gives an exact non-perturbative treatment of a periodically-driven system in terms of an infinite dimensional time-independent matrix. Through adiabatic elimination of high-energy states of this matrix, an effective Hamiltonian can be derived, which gives an accurate description of the system on slow time-scales in terms of low-energy states. For the adiabatic elimination of high-energy states, we apply a generalized approach to Salwen's nearly degenerate perturbation theory \cite{Salwen55}.

\subsection{Single-mode Floquet theory}
\label{ssec:single_mode_Floquet_theory}
We assume that $H(t)$ is a T-periodic Hamiltonian defined on a Hilbert space $\mathcal{H}$ such that 
\begin{equation} \label{eq:hamiltonian-general}
	H(t)  = H(t+T) = H_0 + H_1(t)
\end{equation}
Here, $H_0$ is the static and non-interacting part of $H(t)$ with eigenstates $\ket{\psi_\alpha}$ and eigenenergies $E^{(0)}_{\alpha}$ and $H_1(t)$ is the interacting and time periodic part. $H_1(t)$ thus contains both, the couplings between the qubits as well as all terms that describe time modulations by an external drive. For two qubits, $\mathcal{H}$ is four dimensional, i.e. $\alpha = 1,2,3,4$. Floquet theory \cite{Floquet1883} states that the solution to the Schr\"odinger equation,
 $i \partial_t \ket{\psi(t)} = H(t) \ket{\psi(t)}$
is given by a T-periodic function, $\ket{\tilde{\psi}(t)} = \ket{\tilde{\psi}(t+T)}$, modulated by a complex prefactor, such that $\ket{\psi(t)} = \exp(-i \epsilon t) \ket{\tilde{\psi}(t)}$. The Schr\"odinger equation is then recast into the form, 
\begin{equation} \label{eq:schrodinger_equation}
 [H(t) - i \partial_t]\ket{\tilde{\psi}(t)} = \epsilon \ket{\tilde{\psi}(t)}, 
\end{equation}
where $\epsilon$ is called the quasienergy of the system and plays the same role as energy in static systems. 
By expressing $H(t)$ and $\tilde{\psi}(t)$ in terms of Fourier components, Shirley \cite{Shirley65} showed that Eq.~(\ref{eq:schrodinger_equation}) can be equivalently written in terms of an infinite-dimensional time-independent Floquet matrix $H_F$ as,
\begin{equation}
H_F \ket{\tilde{\psi} \rangle} =  \epsilon \ket{\tilde{\psi} \rangle}  
\end{equation}
which is defined over a composite Hilbert space $\mathcal{H}_c = \mathcal{H} \otimes \mathcal{T}$ \cite{Sambe72}, where $\mathcal{T}$ is the Fourier space of T-periodic functions. The basis of $\mathcal{T}$ is denoted by $\{ \ket{m} \}$, $m \in \mathbb{Z}$ with the property that $\braket{t|m} =\exp{(i m \omega_d t)}$ and $\omega_d = 2 \pi/T$. $m$ is the Fourier index and has the physical meaning of the number of quanta (photons) exchanged with the drive. The states $\ket{ \psi_{\alpha}, m \rangle} = \ket{\psi_{\alpha}} \otimes \ket{m}$ form a basis of $\mathcal{H}_c$, the quasienergies $\epsilon$ are the eigenvalues of the Floquet matrix $H_F$, whose matrix elements read,
\begin{equation}
 \bra{\langle \psi_{\beta}, n} H_F \ket{\psi_{\alpha}, m \rangle} = \tilde{H}^{(n-m)}_{\beta \alpha } + n \omega_d \delta_{\beta \alpha} \delta_{nm}
\end{equation}
where $\tilde{H}^{(n)}$ are the Fourier components of $H(t)$, i.e. $H(t) = \sum_n \tilde{H}^{(n)} e^{ i  n \omega_d t}$. The construction of the Floquet matrix is facillitated by introducing operators $F_n$ and $N$ in the Fourier space $\mathcal{T}$,
\begin{equation} \label{eq:fourier_op}
\braket{m' |F_n| m} = \delta_{m'-m,n}, \quad \braket{m'|N|m} = m  \delta_{m',m},
\end{equation}
so that
\begin{equation} \label{eq:Floquet_Hamiltonian}
 H_F = \sum_{n} \tilde{H}^{(n)} \otimes F_{n} + \omega_d \mathbb{1} \otimes  N
\end{equation}
The explicit form of the Floquet matrix $H_F$ is given in appendix \ref{sec:explicit-Floquet}.

One can decompose $H_F$ into a non-interacting part $H_{F0}$ (corresponding to $H_0$) and an interaction part $V$ (corresponding to $H_1$) such that $H_{F} = H_{F0} + V$. Therefore the states $\ket{ \psi_{\alpha}, m \rangle}$ are the eigenstates of $H_{F0}$ with quasienergy $\epsilon^{(0)}_{(\alpha,m)} = E^{(0)}_{\alpha} + m \omega_d$. We denote the eigenstates of the full Floquet matrix $H_F$ corresponding to the bare states $\ket{ \psi_{\alpha}, m \rangle}$ by $\ket{ \tilde{\psi}_{\alpha, m} \rangle}$ and their quasienergies by $\epsilon_{\alpha, m}$. These quasienergies $\epsilon_{(\alpha,m)}$ satisfy the periodicity relation, $ \epsilon_{(\alpha, m)} = \epsilon_{(\alpha ,0)} + m \omega$.

For the Floquet matrix $H_F$, we can define a time-evolution operator $U_F(t-t_0)$ from an initial time $t_0$ to a final time $t$ as,
\begin{equation}
 U_F(t-t_0) = e^{-i H_F (t-t_0)}, 
\end{equation}
which is related to the time-evolution operator in the Hilbert space, $U(t,t_0)$, via
\begin{equation} \label{eq:time_evolution_pictures2}
 \braket{\psi_{\beta} | U(t,t_0) |\psi_{\alpha}} = \sum_m \braket{\langle\psi_{\beta},m|U_F(t-t_0)|\psi_{\alpha},0\rangle} e^{i m \omega t}
\end{equation}
The essence of our Floquet engineering approach is that, by choosing an appropriate working point for the modulation, we can find a set of four Floquet eigenstates of $H_{F0}$ that describe two coupled spin-1/2 systems and form a four-fold degenerate manifold $\mathcal{S}$, which oscillates at zero frequency in a suitable rotating frame. 
By virtue of Eq.~(\ref{eq:time_evolution_pictures2}) the Floquet time-evolution in this rotating frame in Floquet space $\mathcal{H}_c$ is identical to the time-evolution in the rotating frame of the qubits in the original Hilbert space $\mathcal{H}$.

The dynamics of the two coupled spins is, in this rotating frame generated by the interaction $V$. The degenerate slow subspace is separated from higher energy state by an energy gap. Hence if the interaction $V$ is weak compared to the energy gap, the dynamics of the Floquet Hamiltonian is confined to the slow space $\mathcal{S}$ and one can adiabatically eliminate higher energy states (fast space) to approximate the dynamics by an effective Hamiltonian that only acts on $\mathcal{S}$.

To test whether the desired effective Hamiltonian and hence the desired evolution is indeed generated, we here analyze the time-dependent transition probability from an initial state $\ket{\psi_{\alpha}}$ to final state $\ket{\psi_{\beta}}$,
\begin{align} \label{eq:time_dependent_transition_probability}
 P_{\alpha \rightarrow \beta}(t-t_0) = & |\braket{\psi_{\beta} | U(t,t_0) |\psi_{\alpha}} |^2 \\
 = &\sum_{n} |\braket{\langle \psi_{\beta}, n|U_F(t-t_0)|\psi_{\alpha}, 0 \rangle}|^2, \nn
\end{align}
which can be time-averaged over $t-t_0$ to give,
\begin{equation} \label{eq:time_independent_transition_probability}
 \bar{P}_{\alpha \rightarrow \beta} =  \sum_{n,\gamma,m} |\braket{\langle \psi_{\beta}, n | \tilde{\psi}_{\gamma,m }\rangle}\braket{\langle \tilde{\psi}_{\gamma, m}|\psi_{\alpha}, 0\rangle}|^2
\end{equation}
We now turn to explain the procedure of adiabatically eliminating fast rotating subspaces of the Floquet Hamiltonian.

\subsection{Adiabatic elimination and derivation of an effective Hamiltonian}
\label{ssec:adiabatic_elimination}
We apply a generalized approach to (nearly) degenerate perturbation theory due to Salwen \cite{Salwen55} to derive the desired effective Hamiltonian. This perturbation theory leads to a matrix equation for the slow space, which should be solved in a self-consistent way to give the quasienergies and eigenstates of the effective Hamiltonian, see appendix \ref{appendix:AppB} for details. 

For brevity we drop index $m$ from $\ket{ \psi_{\alpha}, m \rangle}$ for the four states in the slow manifold and denote them by $\ket{\psi_{\alpha} \rangle} \equiv \ket{\psi_{\alpha},m \rangle}$, assuming that $\alpha$ tacitly distinguishes $m$. Hence $\ket{\psi_{\alpha} }$ is a state in $\mathcal{H}$ and $\ket{\psi_{\alpha} \rangle}$ is a state in the slow manifold of the composite space $\mathcal{H}_c$. Accordingly the quasienergy corresponding to $\ket{\psi_{\alpha} \rangle}$ is denoted by $\epsilon^{(0)}_{\alpha} \equiv \epsilon^{(0)}_{(\alpha,m)}$. For the Floquet matrix $H_{F}$, the generalized Salwen perturbation theory leads to the matrix equation,
\begin{align} \label{eq:slow_space_equation}
& \sum_{\beta=1}^4 h_{\alpha\beta}(\epsilon)  \braket{\langle \psi_{\beta} | \tilde{\psi} \rangle} = \epsilon \braket{\langle \psi_{\alpha} | \tilde{\psi} \rangle} \quad \text{with }\\
 & \quad h_{\alpha\beta}(\epsilon) = \braket{\langle \psi_{\alpha} | H_{F0} + T(\epsilon) | \psi_{\beta} \rangle}\nn
\end{align}
for $\alpha = 1,2,3,4$, the solution of which determines the quasienergies $\epsilon_{\alpha}$ and Floquet states $\ket{\tilde{\psi}_{\alpha} \rangle}$ of the slow manifold of the interacting system. Here,
\begin{equation} \label{eq:scattering_marix}
 T(\epsilon)  = \sum_{k=0}^{+\infty} [V G_{\mathcal{P}}]^k V
\end{equation}
is the scattering matrix, where $G_{\mathcal{P}}  = \sum_{\alpha,m} ' (\ket{ \psi_{\alpha}, m \rangle} \bra{\langle \psi_{\alpha}, m } ) /[\epsilon - \epsilon^{(0)}_{(\alpha,m)} ]$ is the Greens function associated with fast space of $H_{F0}$ and the notation $\sum '$ indicates that the degenerate space is excluded from the sum. Since $T$ is a function of $\epsilon$, the set of equations (\ref{eq:slow_space_equation}) needs to be solved self-consistently.

In the absence of interactions, $T(\epsilon) \equiv 0$ and the Floquet quasienergies $\epsilon_j= \epsilon_{j}^{(0)}$ for $j=1,2,3,4$ are recovered from Eq.~(\ref{eq:slow_space_equation}). For nonzero interactions, we make the ansatz that the quasienergies $\epsilon_{\alpha}$, and thus $T(\epsilon)$ and $h_{\alpha \beta}$, associated to the slow space can be written as a perturbative expansion in interaction strength $\lambda$ ($V \propto \lambda$),
\begin{eqnarray}
	\epsilon_{\alpha} & = & \epsilon_{\alpha}^{(0)} + \delta \epsilon_{\alpha}, \quad \textrm{with} \quad
\delta \epsilon_{\alpha} = \sum_{p=1}^{\infty} \kappa_{\alpha,p} \lambda^p \label{eq:quasienergy_expansion}\\
    h_{\alpha \beta} & = & h_{\alpha \beta}(\epsilon_{\alpha}^{(0)})  + \sum_{p=1} ^{\infty} h_{\alpha \beta,p} \lambda^p \label{eq:matrix_elements}
\end{eqnarray}
where $\kappa_{\alpha,p}$ and $h_{\alpha \beta,p}$ are the expansion coefficients.
In practice we truncate the expansion at a desired cut-off order $p = p_{c}$ and calculate the eigenvalues of the effective Hamiltonian using Eq.~(\ref{eq:matrix_elements}). Note that the $p$-th order of $\epsilon_{\alpha}$ modifies the matrix elements $h_{\alpha \beta}$ in $(p+2)$-th order because of the form of $T(\epsilon)$ in Eq.~(\ref{eq:scattering_marix}), which subsequently changes the quasienergies of the updated matrix in $(p+2)$-th order. Hence the results (quasienergies) are exact up to $p$-th order. This suggests that the $\kappa_{\alpha,p}$ are determined by the following procedure: 1) truncate $\epsilon_{\alpha}$ to $p_c$-th order, 2) calculate the matrix elements and quasienergies of the matrix and 3) compare the expansion (\ref{eq:quasienergy_expansion}) and calculated quasienergies.

Since $V$ contains both, couplings between qubits and external drives, the expansions in Eqs.~(\ref{eq:quasienergy_expansion}) and (\ref{eq:matrix_elements}) expand in the strengths of the qubit-qubit couplings as well as in the amplitude of the drives. We consider scenarios, where both processes occur at similar rates, which justifies treating them on the same footing in our perturbation theory.

\subsection{Generalization to many-mode Floquet-theory}
\label{ssec:many_mode_Floquet}
The single-mode Floquet theory can be generalized to a case where the system is not strictly periodic but rather quasi-periodic, i.e. it is driven by a multi-tone drive \cite{Ho83}. Here we present the theory for a bimodal driven system which we apply in the subsequent sections. The generalization to higher number of modes is straight forward. 

We consider a static system $H_0$ as in Sec. \ref{ssec:single_mode_Floquet_theory}, but now assume the system is simultaneously modulated by two periodic perturbations $H_1(t)$ and $H_2(t)$ with frequencies $\omega_{d1}$ and $\omega_{d2}$. For this bimodal driven system $H(t) = H_0 + \sum_{i=1,2} H_i (t)$, there is a time-independent Floquet representation in the composite Hilbert space $\mathcal{H}_c = \mathcal{H} \otimes \mathcal{T}_1 \otimes \mathcal{T}_2$, where $\mathcal{T}_i$ is defined to be the Fourier space of functions with periodicity $2\pi/{\omega_{di}}$ for $i=1,2$. The basis for the composite Hilbert space is denoted by $\ket{\psi_{\alpha},m_1,m_2\rangle}\equiv\ket{\psi_{\alpha}}\otimes\ket{m_1}\otimes\ket{m_2}$, where $\{\ket{m_i} | m_i \in \mathbb{Z}\}$ form a basis for $\mathcal{T}_i$  for $i = 1,2$. The Hamiltonian can be written in terms of its Fourier components as,
\begin{equation} \label{eq:many_mode_Fourier_decomposition}
 H = \sum_{(n_1,n_2)} \tilde{H}^{(n_1,n_2)} \exp{[i (n_1\omega_{d1} + n_2 \omega_{d2}) t]}
\end{equation}
and the Floquet matrix reads,
\begin{align}\label{eq:many_mode_Floquet_Hamiltonian}
	H_F = & \sum_{(n_1,n_2)} \tilde{H}^{(n_1,n_2)}\otimes F_{n_1} \otimes F_{n_2} \\
	+ & \omega_{d1} \, \mathbb{1}  \otimes \mathbb{1} \otimes N + \omega_{d2} \, \mathbb{1} \otimes N \otimes \mathbb{1} \nn
\end{align}
Following Eq.~(\ref{eq:hamiltonian-general}), we decompose $H_F$ into non-interacting part $H_{F0}$, corresponding to $H_0$, and an interaction part $V$, corresponding to $H_1$ and $H_2$. The states $\ket{\psi_{\alpha},m_1,m_2\rangle}$ are thus the eigenstates of $H_{F0}$ with quasienergies $\epsilon^{(0)}_{(\alpha,m_1,m_2)} = E^{(0)}_{\alpha} + m_1 \omega_{d1} + m_2 \omega_{d2}$ and we denote the eigenstates and quasienergies of the interacting system $H_F$ by $\ket{\tilde{\psi}_{\alpha},m_1,m_2\rangle}$ and $\epsilon_{(\alpha,m_1,m_2)}$ --- with periodicity relation $ \epsilon_{(\alpha,m_1,m_2)} = \epsilon_{(\alpha,0,0)} + m_1 \omega_{d1} + m_2 \omega_{d2}$. Here, the projection relations for the time-evolution operators from the composite space to the Hilbert space read,
\begin{equation}
\begin{split}
& \braket{\psi_{\beta} | U(t,t_0) |\psi_{\alpha}} = \\
& \hspace{6mm} = \sum_{m_1,m_2} \braket{\langle\psi_{\beta},m_1,m_2|U_F(t-t_0)|\psi_{\alpha},0,0\rangle} \times \\
& \hspace{28mm} \times \exp\left[i (m_1 \omega_{d1}  + m_2 \omega_{d2}) t \right]
\end{split}
\end{equation}
Accordingly one can define time-dependent and time-averaged transition probabilities as in the case of single-mode system, see Eqs. (\ref{eq:time_dependent_transition_probability}) and (\ref{eq:time_independent_transition_probability}). 

Note that the definition for the composite space and the form of the many-mode Fourier decomposition Eq.~(\ref{eq:many_mode_Fourier_decomposition}) is not unique here. Depending on the explicit form of $H_i(t)$, one may find it more convinient to redefine the Fourier spaces $\mathcal{T}_i$. For example if $\omega_{d1} = \omega_1 + \omega_2$ and $\omega_{d2} = \omega_1 - \omega_2$, we could equivalently work in the Fourier space of the functions with periodicity $2 \pi/{\omega_i}$. 

To find the effective interaction Hamiltonians, we proceed as in the single mode case and adiabatically eliminate all Floquet states except for the four in the low-energy manifold to effectively describe the infinite-dimensional Floquet matrix by a $4 \times 4$~matrix. Dropping again the Fourier index for the low-energy states and quasienergies and denoting them by $\ket{\psi_{\alpha} \rangle}$ and $\epsilon^{(0)}_{\alpha}$, the Green's function of the fast space reads $G_{\mathcal{P}}(\epsilon) = \sum_{\alpha,m_1,m_2}^{'}  (\ket{ \psi_{\alpha}, m_1,m_2 \rangle} \bra{\langle \psi_{\alpha}, m_1,m_2 })/(\epsilon - \epsilon^{(0)}_{(\alpha,m_1,m_2)})$ where $\epsilon$ is the quasienergy of the degenerate manifold in the interacting limit and the sum runs over the fast space where the degenerate states are excluded. The procedure of constructing the self-consistent matrix equation and solving for its quasienergies runs analogous to the single-mode case, c.f. Eqs. (\ref{eq:slow_space_equation}), (\ref{eq:scattering_marix}), (\ref{eq:quasienergy_expansion}) and (\ref{eq:matrix_elements}). 

We now turn to apply the formalism developed above to specific physical set-ups and derive the corresponding effective interaction Hamiltonians.

\section{Driven qubit schemes}
\label{sec:driven_qubit_schemes}
\begin{figure}
 \centering
 \includegraphics[width=0.4\textwidth]{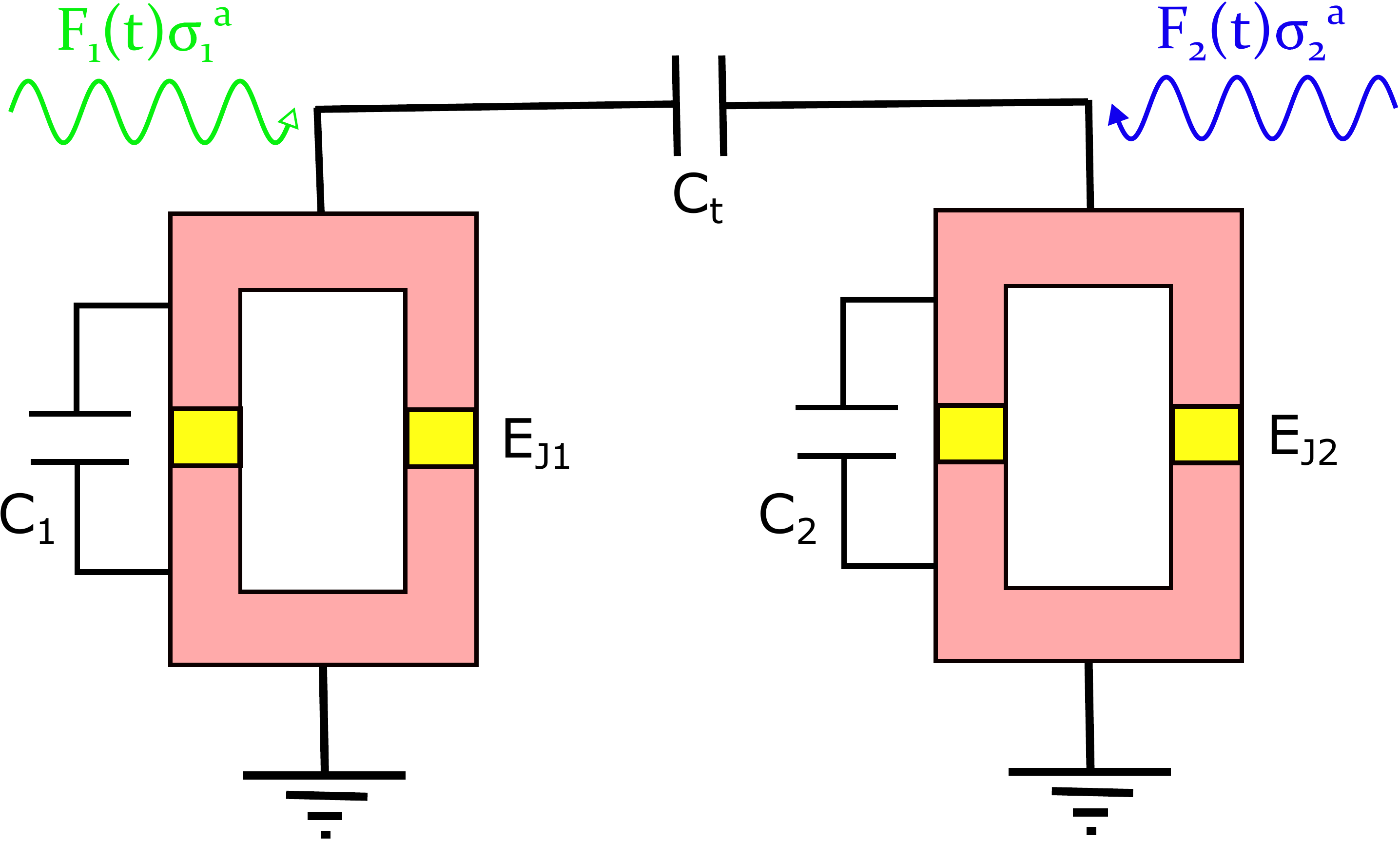}
 \caption{\label{fig:circuit} Driven qubit scheme. We here consider two capacitively coupled transmon qubits driven by single mode or bimodal driving functions. In the single mode case, one of the qubits (e.g. qubit 1) is modulated by a drive as defined in Eq.~(\ref{eq:single_mode}). In this system the $ZX$, $ZY$, hopping and squeezing terms are engineered. In the bimodal case, both of the qubits are driven via two-mode Hamiltonian (\ref{eq:bimodal_mod}). In this case $XX$, $YY$, $XY$, $YX$ and $ZZ$ spin interactions are enabled.}
\end{figure}

In this section we apply the formalism developed in Sec. \ref{sec:theoretical-toolbox} to capacitively coupled qubits with single-mode or bimodal driving.
We consider two superconducting Transmon qubits \cite{Koch07} with Josephson energies $E_{Ji}$ and capacitances $C_{i}$ ($i = 1,2 $), see Fig.~\ref{fig:circuit}, that are coupled via a capacitance $C_{t}$ with $C_t \ll C_j$.
The dynamics of the system can be described in terms of the nodal phases $\phi_1$ ($\phi_2$) and their conjugate variables $n_1$ ($n_2$) with the Hamiltonian,
\begin{equation} \label{eq:Hamiltonian}
\overline{H} = \sum_{j=1,2} -E_{Jj}  \cos(\phi_j) + E_{Cj} n_j^2 + E_{cc} n_1 n_2, 
\end{equation}
where $E_{Cj} = \hbar^2(2 \phi_0^2 C_j)^{-1}(1-C_t/C_j)$, $E_{cc} = \hbar^2(\phi_0^2 C_1C_2)^{-1} C_t$ and $\phi_0 = \hbar/(2 e)$ is the rescaled flux quantum.
In the phase regime ($E_{Jj}/E_{Cj} \gg 1$), relevant for Transmon qubits, we have $||\phi_j|| \ll 1$ and can approximate $\cos(\phi_j) \approx  - \phi_j^2/2! + \phi_j^4/24! + \text{const} $. The Hamiltonian in this regime describes two nonlinear oscillators that are coupled through the last term in Eq.~(\ref{eq:Hamiltonian}). It is quantized via the canonical commutation relations $[\phi_i,n_j] = i  \delta_{ij}$, resulting in the substitutions $\phi_j = \overline{\phi}_j (a_j +a_j^{\dagger})$ and $n_j = -i (2\overline{\phi}_j)^{-1} (a_j - a_j^{\dagger})$, where $a^{\dagger}_j (a_j)$ are creation (destruction) operators and $\overline{\phi}_{j} = (E_{Cj}/2E_{Jj})^{\frac{1}{4}}$ the zero-point motion amplitude of qubit $j$.
The quantized Hamiltonian reads  ($\hbar = 1$),
\begin{equation} \label{eq:hamiltonian_capacitive_coupling}
 \overline{H} = \sum_{j=1,2} (\omega_j a_j^{\dagger} a_j  - U_j a^{\dagger}_j a^{\dagger}_j a_ja_j) + g_c (a_1- a_1^{\dagger})(a_2 -a_2^{\dagger})  
\end{equation}
with,
\begin{equation}
 \begin{split}
  & \omega_j =  \sqrt{2 E_{Jj} E_{Cj}} + 2U_j, \quad j \in \{ 1,2 \} \\
  & U_j = -E_{Jj} \overline{\phi}_{j}^4/4 , \\
  & g_c = -E_{cc} / (4\overline{\phi}_1 \overline{\phi}_2) 
 \end{split}
\end{equation}
Throughout the text, we assume that the two qubits are detuned from each other with a detuning  $\Delta = \omega_1 - \omega_2>0$ of the order of $\omega_j$ ($j=1,2$). 
Due to their nonlinear spectrum, we further assume that the Hamiltonian can be truncated to the single excitation subspace, $\ket{0_j}$ and $\ket{1_j}$ with $a^{\dagger}_j a_j \ket{n_j} = n_j \ket{n_j}$, where it reads in terms of Pauli operators,
\begin{equation} \label{eq:H0}
 \overline{H} = \sum_{j=1,2} \omega_j \sigma_j^{+} \sigma_j^- + g_c \sigma_1^y \sigma_2^y. 
\end{equation}
Since we consider the coupling $g_c$ in a perturbation expansion, the background Hamiltonian $H_0$, c.f. Eq.~(\ref{eq:hamiltonian-general}, reads,
\begin{equation} \label{eq:ham_background}
	H_0 = \sum_{j=1,2} \omega_j \sigma_j^{+} \sigma_j^-.
\end{equation}
We introduce the states $\ket{\psi_1} = \ket{1_1,1_2}$, $\ket{\psi_2} = \ket{1_1,0_2}$, $\ket{\psi_3} = \ket{0_1,1_2}$ and $\ket{\psi_4} = \ket{0_1,0_2}$, which form a basis for the Hilbert space. They are the eigenstates of $H_0$ with eigenenergies $E_{1}^{(0)} = \omega_1 + \omega_2$, $E_{2}^{(0)} = \omega_1$, $E_{3}^{(0)} = \omega_2$ and $E_{4}^{(0)} = 0$.
In the limit where $g_c \ll \omega_j$, the interaction between the two qubits is ineffective due to their detuning.
For this system single mode and bimodal driving are implemented as follows. 

By a single-mode drive we denote the scenario, where one of the qubits (e.g. qubit 1) is modulated by the periodic Hamiltonian,
\begin{equation} \label{eq:single_mode}
 H_d(t) = F_1(t) \sigma^{a}_{1} \quad \text{for} \quad  a = x, z, \\
\end{equation}
where
\begin{equation}
	F_1(t) = b \cos(\omega_{d} t + \theta)
\end{equation}
and $b$ is the strength, $\theta$ the phase, and $\omega_d$ the frequency of the modulation. We refer to the two cases with $a=x$ and $a=z$ as {\it transverse} and {\it longitudinal modulations}. In an implementation, the transverse modulation is realized by coupling the first qubit to a charge line which is typically used for the standard readout processes. For longitudinal modulation, a small oscillating flux is threaded on top of the dc flux through the loop of the qubit's SQUID to modulate its effective Josephson energy, c.f. \cite{Sameti16}.   

As a bimodal modulation we refer to a quasiperiodic Hamiltonian,
\begin{equation} \label{eq:bimodal_mod}
 H_d(t) = F_1 \sigma_1^{a} + F_2 \sigma_2^{a} \quad \text{for} \quad  a = x, z, \\
 \end{equation}
where,
\begin{equation} \label{eq:driving_function}
 F_i = \sum_{j=1,2} b_{ij} \cos(\omega_{dj} t + \theta_{ij}) \quad \text{for} \quad i = 1,2
\end{equation}
and $b_{ij}$ are the amplitudes of the modulations. $\omega_{d1}$ and $\omega_{d2}$ are the two {\it incommensurate} frequencies, and $\theta_{ij}$ possible relative phases of the drives.

In all cases considered here, we assume each $\omega_{dj}$ is off-resonant with respect to the transition frequency $\omega_j$ of the corresponding driven qubit(s). This ensures that the drive cannot create excitations in the qubit it drives without a coupling to another qubit.
Moreover, for both driving scenarios, the total Hamiltonian can be split as in Eq.~(\ref{eq:hamiltonian-general}), where
\begin{equation}
	H_1(t) =  g_c \sigma_1^y \sigma_2^y + H_d(t)
\end{equation}
with $H_d(t)$ given by Eqs. (\ref{eq:single_mode}) or (\ref{eq:bimodal_mod}) and $H_0$ as in Eq.~(\ref{eq:ham_background}).

In the following sections, we derive an effective time-independent Hamiltonian in the rotating frame of the qubits by means of the Floquet formalism. We will show that arbitrary spin-spin interaction can be realized in this system by choosing the right driving scheme. We first summarize our results for this part. 

\subsection{Summary of results}
A large variety of spin-spin interaction can be realized in this system by choosing the right driving scheme. In the case of a single-mode drive, these are:
\begin{enumerate}
	\item When the first qubit is driven with $a=x$ and $\omega_d = \omega_2$, the interaction $J_{zx} \sigma_1^z \sigma_2^x$ and $J_{zy} \sigma_1^z \sigma_2^y$ are implemented for $\theta=0$ and $\theta=\pi$. This interaction is second order including single photon exchange with the drive i.e. $J_{zx} \sim J_{zy} \sim  \frac{2 \omega_2}{\omega_1^2-\omega_2^2} g_c b$. Alternatively by driving the second qubit one can realize $\sigma_1^x \sigma_2^z$ and $\sigma_1^y \sigma_2^z$.
	\item For $a=z$, $\theta=0$, the hopping term $J_{h}\sigma_1^+ \sigma_2^- + \text{H.c.}$ is enabled if $\omega_d= \omega_1-\omega_2$ and  the squeezing term $J_s\sigma_1^+ \sigma_2^+ + \text{H.c.}$ is enabled if $\omega_d= \omega_1+\omega_2$. Both processes are second order with single photon exchange and hence $J_s \sim \frac{2 b g_c}{\omega_1+\omega_2}$ and $J_h \sim \frac{2b g_c}{\omega_1-\omega_2} $.
\end{enumerate}
In the case of bimodal driving with modulation frequencies $\omega_{d1} = \omega_1 + \omega_2$ and $\omega_{d2} = \omega_1 - \omega_2$ the realizable interactions are:
\begin{enumerate}
\setcounter{enumi}{2}
	\item For $a=z$ and $\theta_{ij} = 0$, $J_{xx} \sigma_1^x \sigma^x_2$ and $J_{yy} \sigma^y_1 \sigma^y_2$ can be realized by setting the modulation amplitudes such that $\eta = (b_{11} + b_{21})/(\omega_1+\omega_2)= \pm (b_{12} - b_{22})/(\omega_1-\omega_2)  $ ($-$ for the $XX$ interaction and $+$ for the $YY$ interaction). The interaction is second-order with the strength $J_{xx} = J_{yy} \sim 2\eta g_c$.
	\item If $a=z$ and $\theta_{ij} = \pi$, we get the interactions $J_{xy} \sigma_1^x \sigma^y_2$ and $J_{yx} \sigma^y_1 \sigma^x_2$ when the modulation amplitudes are chosen such that $\eta =  (b_{11} + b_{21})/(\omega_1+\omega_2)=\pm (b_{12} - b_{22})/(\omega_1-\omega_2) $ ($-$ for $xy$ and $+$ for $yx$).  The interaction strength is $J_{xy} = J_{yx} \sim 2\eta g_c$.
	\item For $a=x$ and $\theta=0$, a $\sigma^z\sigma^z$ interaction is implemented, which is a third order term in the coupling strength and thus is the weakest among all other types of effective spin-spin interactions (with a single photon exchange).
\end{enumerate}

\subsection{Single-mode driven system}
We first consider a case in which one of the qubits is driven as described in Eq.~(\ref{eq:single_mode}), see Fig.~\ref{fig:circuit}.  We use the equivalent time-independent Floquet representation of the system, c.f. Eq.~(\ref{eq:Floquet_Hamiltonian}), where, for the Hamiltonian (\ref{eq:hamiltonian-general}), there are just three Fourier components,
\begin{equation}
 \tilde{H}^{(0)} = \overline{H} \quad \textrm{and} \quad
 \tilde{H}^{(1)} = (\tilde{H}^{(-1)})^* = (b/2)e^{i \theta} \sigma_1^{a}
\end{equation}
from which we can construct the Floquet matrix $H_F$ according to Eq.~(\ref{eq:Floquet_Hamiltonian}). 
For this model we now first discuss the engineering of interactions at single photon transitions for transverse ($a=x$) and longitudinal ($a=z$) modulations and then discuss the possibility of engineering interactions at multi-photon exchange.

\subsubsection{Transverse modulation at single photon transition} \label{sssec:ax_single}
\begin{figure}
  \centering
  \subfloat[]{\includegraphics[width=80mm]{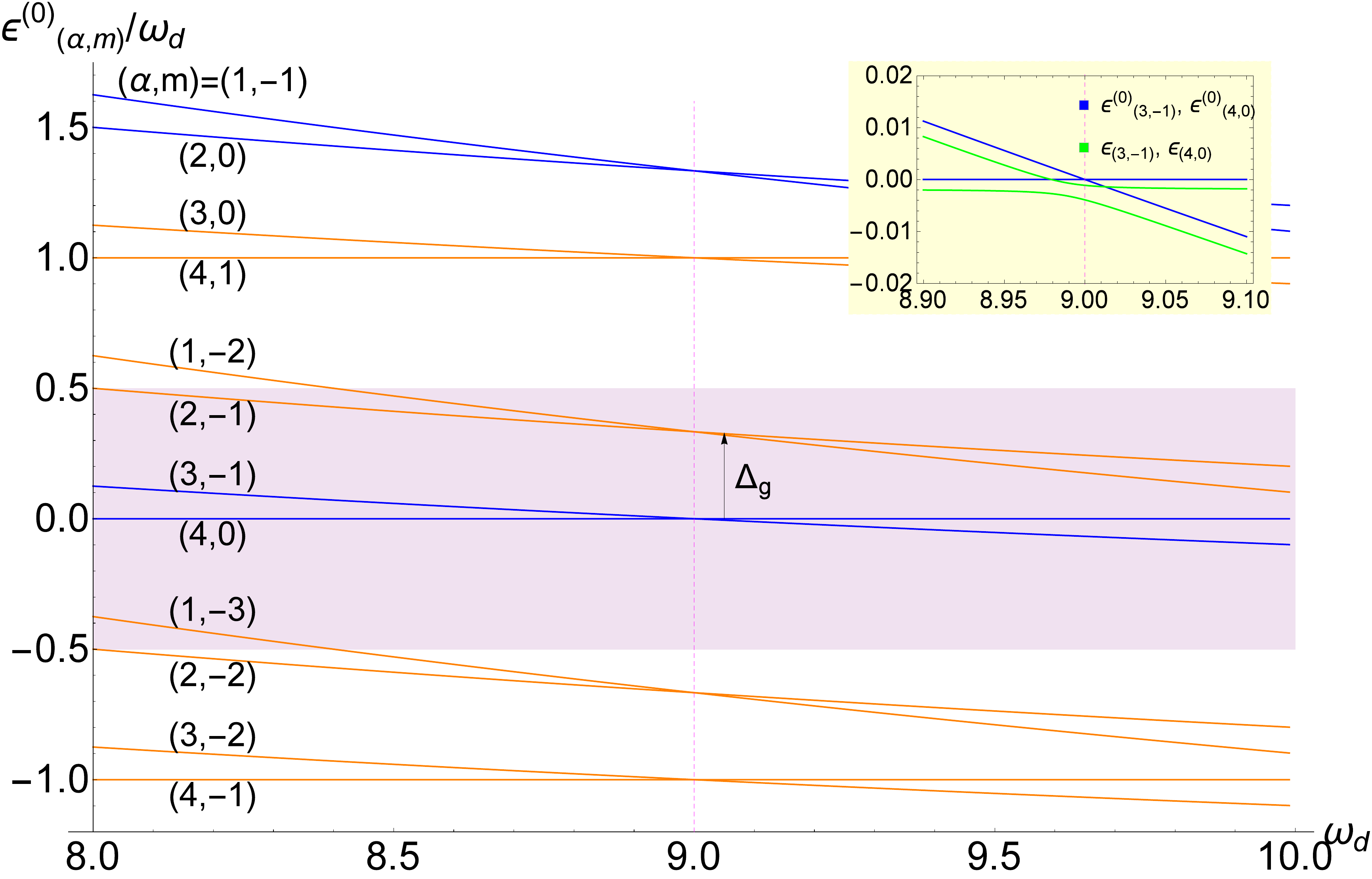}\label{fig:21}}
  \hfill
  \subfloat[]{\includegraphics[width=80mm]{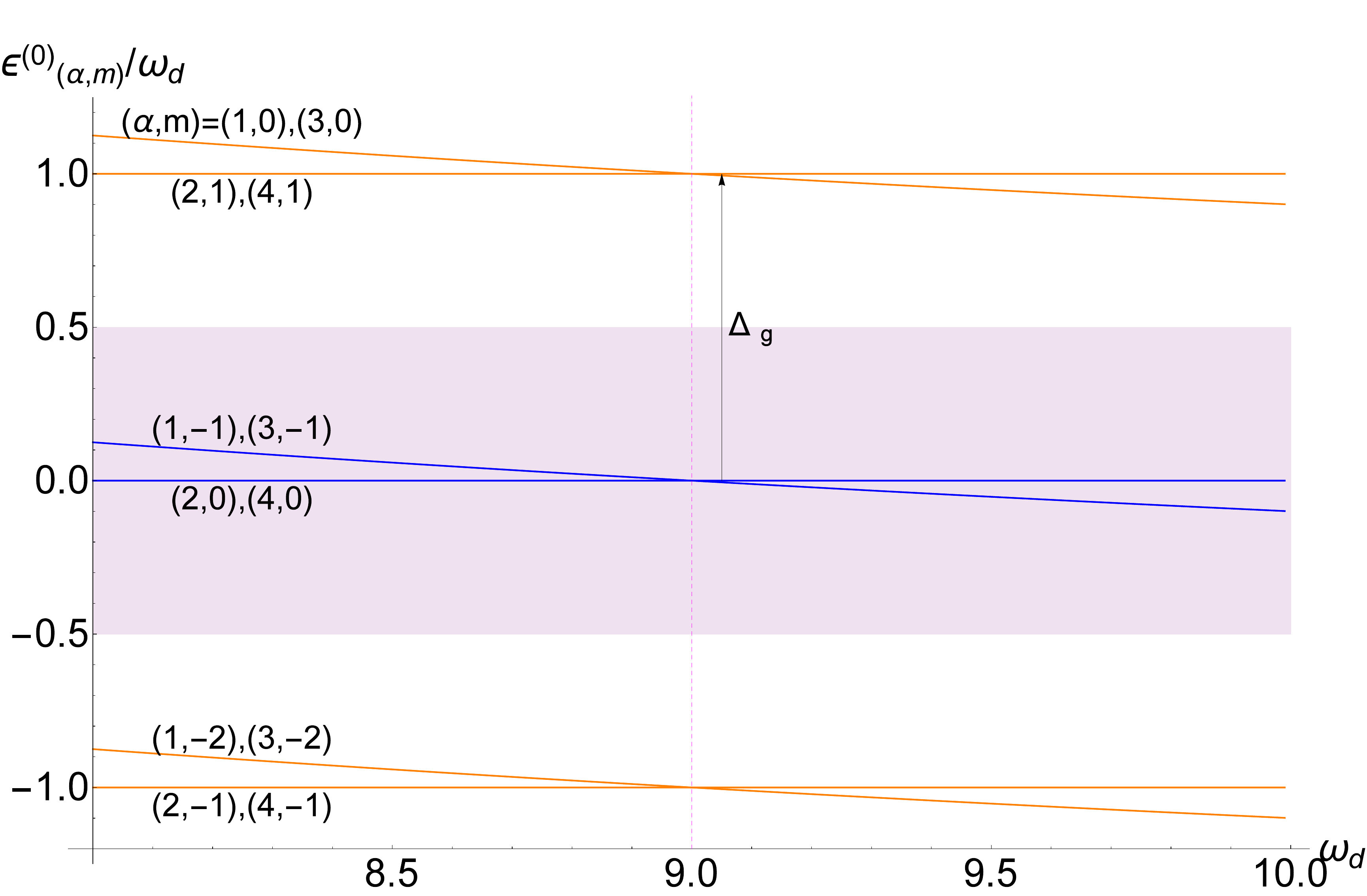}\label{fig:22}}

  \caption{(a) Low-lying quasienergies of the Floquet Matrix $H_{F0}$ as a function of $\omega_d$(GHz) for $\omega_1/(2 \pi) = 12$ GHz, $\omega_2/(2 \pi) = 9$ GHz. The states with blue color give rise to the effective Hamiltonian in the rotating frame of the qubits. $\Delta_g$ is the relevant gap for $H_{F0}$ which is $\Delta_g \approx \omega_1 - \omega_2$ at the crossing point. The shaded area is the Floquet Brillouin zone. The inset shows the bare and dressed quasienergies of degenerate states at zero energy at the crossing point. (b) Quasienergies in the rotating frame defined with respect to $ \omega_1 \sigma_1^+ \sigma_1^- \otimes  \mathbb{1}$. The gap in the rotating frame changes to $\Delta_g \approx \omega_d$ at the crossing point.} 
  \label{fig:One}
\end{figure}
We here explain our Floquet approach for generating an effective parity-breaking interaction $\sigma_1^{z} \sigma_2^{x}$ or $\sigma_1^z\sigma_2^y$ in a single-mode driven system with transverse modulation. This also reproduces existing gate schemes \cite{Rigetti10, Chow11, Reagor18}.

We first assume $\theta=0$. An example of quasienergies of the bare Floquet matrix $H_{F0}$ as a function of $\omega_d$ for $\omega_1/(2 \pi)= 12$ GHz and $\omega_2/(2 \pi) =9$ GHz is shown in Fig.~\ref{fig:One}.
In the frame defined by $\mathcal{U}_{q1} = \exp{(-i \omega_1 t \sigma_1^+ \sigma_1^- \otimes \mathbb{1})}$ in the composite space, the Floquet matrix $H_{F0}$ transforms to $H_{F0} \to H_{F0}  - \omega_1 \sigma_1^+ \sigma_1^-  \otimes \mathbb{1}$. In this frame, which is equivalent to the rotating frame of the qubits, and in the absence of interactions, the Floquet states $\ket{\psi_1, m \rangle}$ and $\ket{\psi_3, m \rangle}$ are degenerate with quasienergy $\omega_2 + m \omega_d$, and the Floquet states $\ket{\psi_2, m \rangle}$ and $\ket{\psi_4, m \rangle}$ are degenerate with quasienergy $m \omega_d$. 

As can be seen in Fig.~\ref{fig:One}, at the specific point $\omega_d = \omega_2$, the four Floquet states $\ket{\psi_1 ,  m =  -1 \rangle}$, $\ket{\psi_2 , m = 0 \rangle}$, $\ket{\psi_3, m = -1 \rangle}$ and $\ket{\psi_4 , m = 0 \rangle}$ form a degenerate manifold $\mathcal{S}$ with quasienergy $\epsilon_{(\alpha , m)}^{(0)} = 0$. Here $m=-1$ indicates that a single-photon exchange with the drive is responsible for the degeneracy of the states. This zero-frequency manifold is separated from states with higher quasienergies by a gap of size $\omega_1 - \omega_2$. We therefore eliminate the high energy states by applying the perturbation toolbox introduced in the Sec. \ref{ssec:adiabatic_elimination}. For this single mode transverse driven system at $\omega_d = \omega_2$, the Salwen matrix has the structure,
\begin{equation} \label{eq:structure_odd}
 \begin{bmatrix}
  h_{11}(\epsilon)   & h_{12}(\epsilon) &0 &0  \\
  h_{21}(\epsilon)   & h_{22}(\epsilon) &0 &0 \\
  0   & 0 &h_{33}(\epsilon) &h_{34}(\epsilon)  \\
  0   & 0 &h_{43}(\epsilon) &h_{44}(\epsilon) \\
 \end{bmatrix}
\end{equation}
where the $h_{\alpha,\beta}$ are as in Eq.~(\ref{eq:slow_space_equation}).
Therefore $\mathcal{S}$ decouples into two subspaces, one spanned by $\mathcal{S}_1 = \{ \ket{ \psi_1 \rangle} , \ket{ \psi_2 \rangle} \}$ and another spanned by $\mathcal{S}_2 = \{ \ket{ \psi_3 \rangle} , \ket{ \psi_4 \rangle} \}$ (As explained in Sec. \ref{sec:theoretical-toolbox} we drop the index $m$ for the degenerate manifold).

Setting $\lambda = b$, c.f. Eq.~(\ref{eq:quasienergy_expansion}), and introducing the ratio $\eta = g_c/b$ we obtain $\delta \epsilon_{\alpha}$ for $\alpha=1, 2$ as given in Eq.~(\ref{eq:quasienergy_expansion}), where the coefficients read,
\begin{align}
	\kappa_{\alpha,1} & = 0 \\
	\kappa_{\alpha,2} & = \frac{-(2+\eta^2)\omega_1 \pm \sqrt{\eta^2 (4+\eta^2)} \omega_2}{\omega_1^2-\omega_2^2}\\
	\kappa_{\alpha,4} & = - \frac{\kappa_{\alpha,2} \tilde\eta (\omega_1^2+\omega_2^2) + 2 \omega_1 (1 \pm \kappa_{\alpha,2} |\eta| \sqrt{2+ \tilde\eta} \omega_2)}{(\omega_1-\omega_2)^2(\omega_1+\omega_2)^2},
\end{align}
where where $\alpha=1(2)$ for the upper(lower) sign, $\tilde\eta = 2+\eta^2$ and we have truncated the expansion at 4th order.
The corresponding eigenstates $\ket{\tilde{\psi}_1 \rangle}$ and $\ket{\tilde{\psi}_2 \rangle}$ are,
\begin{equation}
 \ket{ \tilde{\psi}_{\alpha} \rangle} = \frac{1}{\mathcal{N}} \left(\frac{-1 \pm \sqrt{4 + \eta^2}}{2} \ket{\psi_1 \rangle} + \ket{\psi_2 \rangle}\right),
\end{equation}
where $\alpha=1(2)$ for the upper(lower) sign and $\mathcal{N}$ is a normalization factor. For the subspace $\mathcal{S}_2$ we get,
\begin{equation}
 \delta \epsilon_3 = - \delta \epsilon_2, \quad \delta \epsilon_4 = - \delta \epsilon_1
\end{equation}
and
\begin{equation}
 \ket{ \tilde{\psi}_{\alpha} \rangle} = \frac{1}{\mathcal{N}} \left(\frac{1 \mp \sqrt{4 + \eta^2}}{2} \ket{\psi_3 \rangle} + \ket{\psi_4 \rangle}\right),
\end{equation}
where $\alpha=3(4)$ for the upper(lower) sign.

The effective Hamiltonian in Flouet space, $\tilde{H}_{eff} = \sum_{\alpha} \epsilon_{\alpha} \ket{ \tilde{\psi}_{\alpha} \rangle} \bra{\langle \tilde{\psi}_{\alpha}}$, can then, after some elementary algebra, be written in terms of spin operators as,
\begin{equation} \label{eq:effective_Hamiltonian_single_photon}
 \tilde{H}_{eff} = \left(\frac{\omega_1}{2} + \delta \omega_1\right) \sigma_1^{z}+ \delta \omega_2 \sigma_2^{z} + J_{zx} \sigma_1^{z} \sigma_2^{x}
\end{equation}
This Hamiltonian describes the effective dynamics in the Floquet space, where $\delta \omega_1$ and $\delta \omega_2$ are frequency shifts incurred on the two qubits due to the interaction, and $J_{zx}$ is the strength of the effective interaction between the qubits. These parameters read,
\begin{align}\label{eq:qubit2shift}
	 \frac{\delta \omega_1}{\omega_1} & = \frac{(2+\eta^2)}{\omega_1^2-\omega_2^2} b^2 \\
	 & -  \frac{((2+\eta^2)^2+2)\omega_1^2 + (3(2+\eta^2)^2+10) \omega_2^2}{(\omega_1^2-\omega_2^2)^3}  b^4, \nn \\
     \frac{\delta \omega_2}{\omega_2} & = -\frac{\eta^2}{\omega_1^2-\omega_2^2} b^2 +  \frac{\eta^2(2+\eta^2)\omega_2(3\omega_1^2 + \omega_2^2)}{(\omega_1^2-\omega_2^2)^3}  b^4, \\
     J_{zx} & = \frac{2 \eta \omega_2}{\omega_1^2-\omega_2^2} b^2 -  \frac{2\eta(2+\eta^2)(3\omega_1^2 + \omega_2^2)\omega_2}{(\omega_1^2-\omega_2^2)^3}  b^4
\end{align} 
An example of the parameters and dynamics of the effective Hamiltonian is given in table \ref{table:table1}.

\begin{table}[t]
\centering
 \begin{tabular}{llll}
\hline
\multicolumn{3}{r}{Parameters (MHz)} \\
\cline{2-4}
    & $\delta \omega_1/(2 \pi)$ & $\delta \omega_2/(2\pi)$ & $J_{zx}/(2 \pi)$ \\
\hline
GSM (2nd order)      &  $-40.952$     & $-12.857$   & $-21.428$     \\
GSM (4th order)       &  $-40.191$     & $-12.499$   &$-20.833$       \\
GSM (6th order)       &  $-40.224$     & $-12.513$   &$-20.860$       \\
Exact result       & $-40.222$    & $-12.515$ & $-20.855$     \\
\hline
\end{tabular}
\caption{Numerical values for frequency shifts $\delta \omega_i$ ($i=1,2$) and interaction strength $J_{zx}$ in Eq.~(\ref{eq:effective_Hamiltonian_single_photon}) for $\omega_1/(2 \pi) =  12$ GHz, $\omega_2/(2 \pi)= \omega_d/(2 \pi) =  9$ GHz, $b/(2 \pi)=250$MHz and $\eta=1.2$. GSM means Generalized Slawen method.}
\label{table:table1}
\end{table}

The expression for $J_{zx}$ shows that the effective interaction to the lowest order is second-order in $b$ and is formed by two processes: (i) a single photon exchange with the drive ($\propto b$) and (ii) a $\sigma^{(1)}_y\sigma^{(2)}_y$ interaction ($\propto g_c = \eta b$).

The physical meaning of the above  interactions can be obtained via Eq.~(\ref{eq:time_evolution_pictures2}) that relates the time-evolution as observed in the lab, $U(t,t_0)$, with time-evolution operator in the Floquet space, $U_F(t,t_0)$.
To simplify the right hand side of Eq.~(\ref{eq:time_evolution_pictures2}), we introduce the operators $\mathcal{U}_{q1} = \exp{(i \omega_1 t \sigma^+_1 \sigma^-_1 \otimes \mathbb{1})}$ and $\mathcal{U}_{q2} = \exp{(i \omega_2 t \sigma^+_2 \sigma^-_2 \otimes \mathbb{1}) }$, with $\mathcal{U}_{q2} \ket{ \psi_{\alpha}, m\rangle} = e^{i m \omega_d t} \ket{ \psi_{\alpha}, m\rangle}$ on the degenerate subspace and $\mathcal{U}_{q1}^{\dagger} \mathcal{U}_{q1} = \mathbb{1}$. We can thus write 
\begin{widetext}
\begin{align}
	\braket{\psi_{\beta} | U(t,t_0) |\psi_{\alpha}} & =\sum_{m_1,m_2} \braket{\langle\psi_{\beta},m_2|\mathcal{U}_{q1}^{\dagger} \mathcal{U}_{q1} U_F(t-t_0) \mathcal{U}_{q1}^{\dagger} \mathcal{U}_{q1}|\psi_{\alpha},m_1\rangle} e^{i (m_1-m_2) \omega_d t}\\
	 &=  \braket{\langle\psi_{\beta}|\mathcal{U}_{q1}^{\dagger} \mathcal{U}_{q2}^{\dagger} \mathcal{U}_{q1} U_F(t-t_0) \mathcal{U}_{q1}^{\dagger} \mathcal{U}_{q2} \mathcal{U}_{q1}|\psi_{\alpha} \rangle}, \nn
\end{align}	
\end{widetext}
where we have ignored all states except for the low-energy manifold. Since the $\mathcal{U}_{qi}$ are defined over the infinite-dimensional Floquet space, their truncation to the low-energy space is,
\begin{equation} \label{eq:qubit_rot_frame}
 U_{q1}(t) = \exp{(i \omega_1 t \sigma_1^{z}/2)}, \quad U_{q2}(t) = \exp{(i \omega_2 t \sigma_2^{z}/2)}
\end{equation}
and we can write
\begin{equation} \label{eq:lab_Floquet_link}
 \braket{\psi_{\beta} | U(t,t_0) |\psi_{\alpha}} = \braket{\psi_{\beta} | U^{\dagger}_{q1}  U^{\dagger}_{q2} U_{eff}  U_{q1}  U_{q2} |\psi_{\alpha}}
\end{equation}
The operators $U_{qi}$ simply define the rotating frame of the qubits and thus $\ket{\psi_i^{(r)}} = U_{q1}  U_{q2} \ket{\psi_i}$ is the state $\ket{\psi_i}$ transformed to the rotating frame of the qubits. Hence Eq.~(\ref{eq:lab_Floquet_link}) simply reads $\braket{\psi_{\beta} | U(t,t_0) |\psi_{\alpha}} = \braket{\psi_{\beta}^{(r)} | U_{eff}(t-t_0)  |\psi_{\alpha}^{(r)}}$. This means $U_{eff}$ is the time evolution operator,
\begin{equation}
 U_{eff} = U_{q1} e^{-i \tilde{H}_{eff} (t-t_0)} U_{q1}^{\dagger} = e^{-i H_{eff} (t-t_0)}
\end{equation}
with,
\begin{equation}
 H_{eff} = \delta \omega_1 \sigma_1^{z}+ \delta \omega_2 \sigma_2^{z} + J_{zx} \sigma_1^{z} \sigma_2^{x} 
\end{equation}
the effective Hamiltonian in the rotating frame of the qubits.

\begin{figure*}
  \centering
  \subfloat[]{\includegraphics[height=70mm]{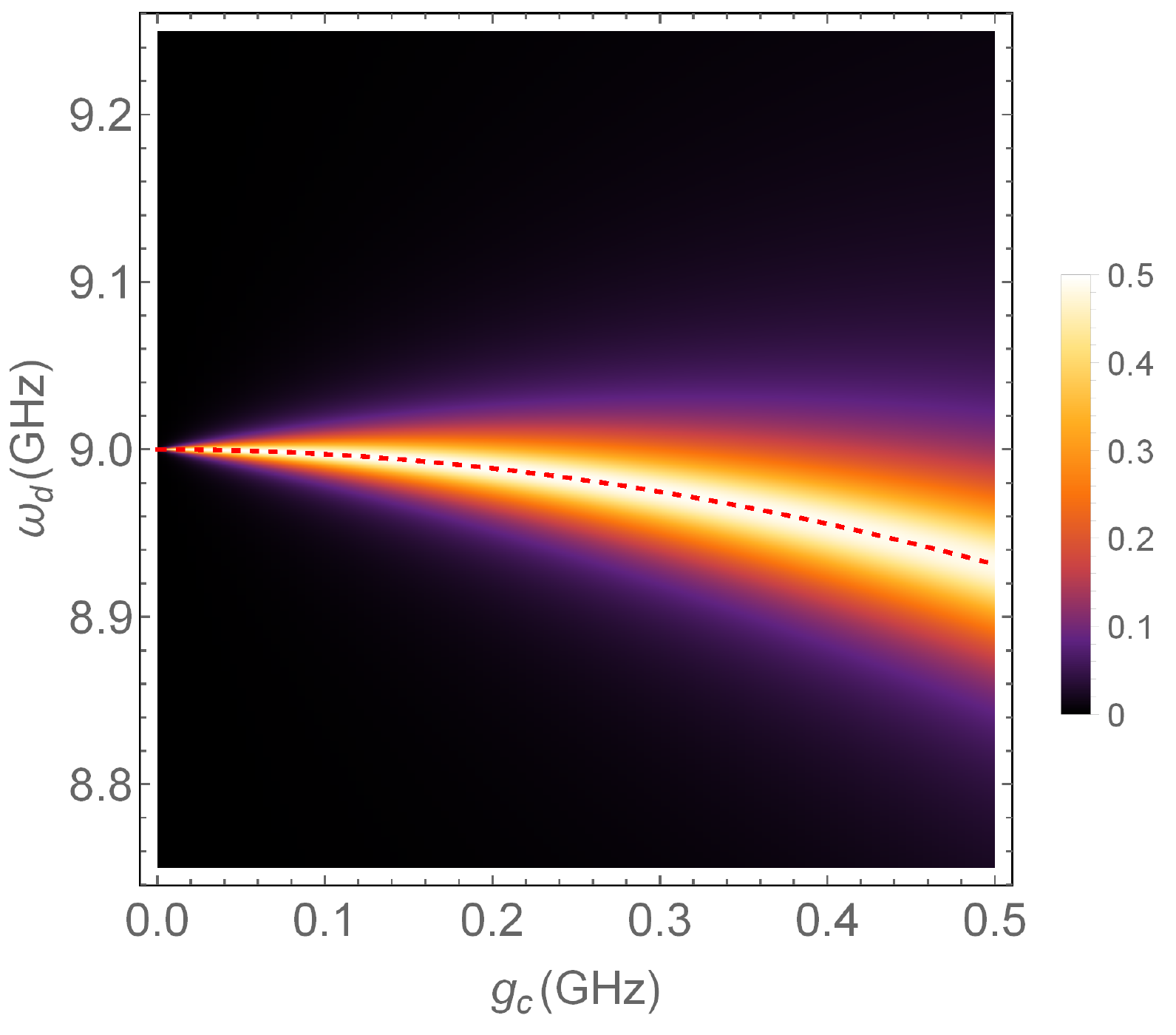}\label{fig:single_photon_resonance1}}
  \hfill
  \subfloat[]{\includegraphics[height=70mm]{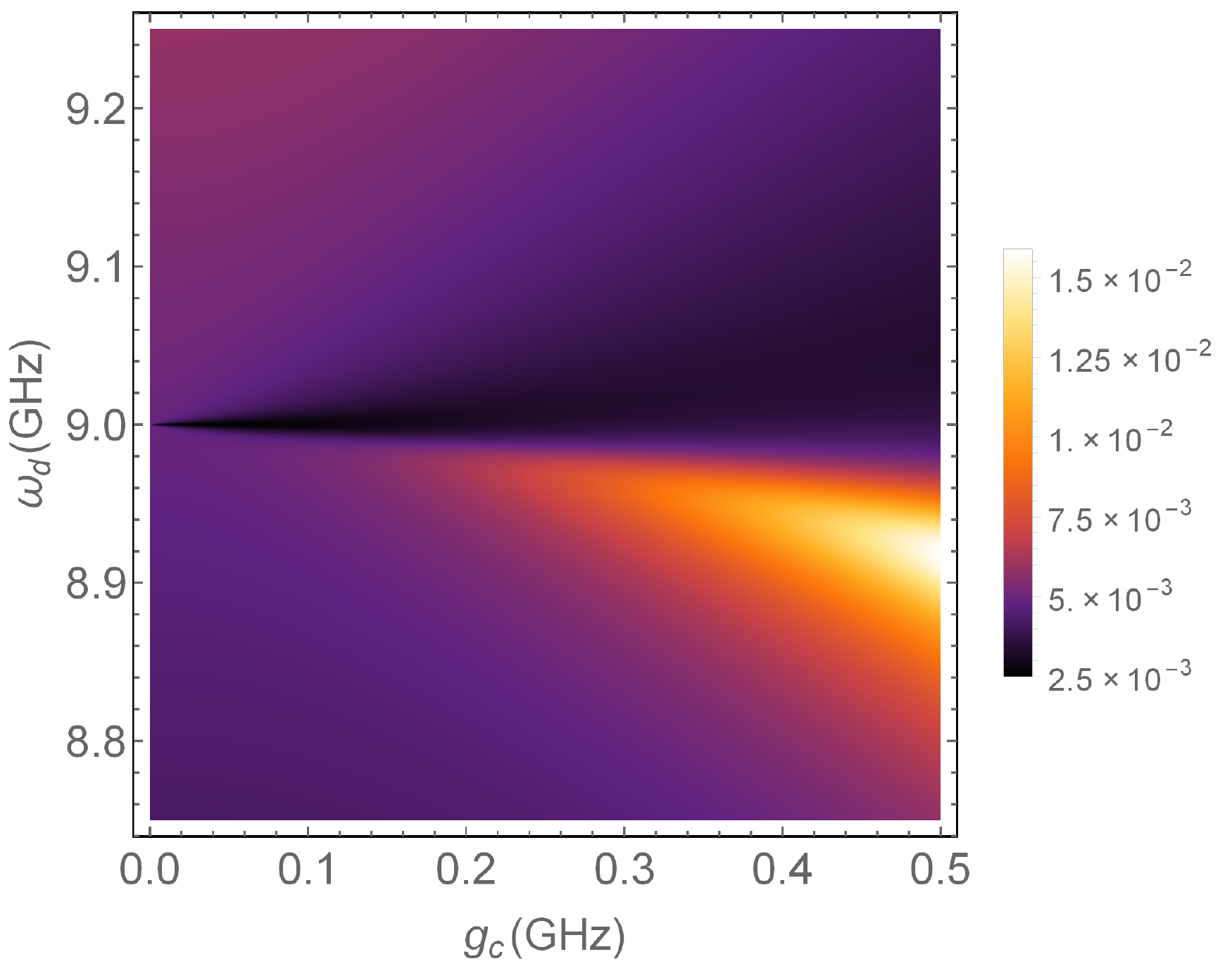}\label{fig:single_photon_resonance2}}
  \caption{ (a) Time-averaged transition probability $\ket{0 0}  \rightarrow \ket{0 1}$ as a function of external drive frequency $\omega_d$ and $g_c$. The resonance is due to the effective interaction $\sigma_z^1 \sigma^y_2$ at single photon transition.   The dashed line is the resonance condition $\omega_d = \omega_2+\delta \omega_2$ in which $\delta \omega_2$ is given by Eq.~(\ref{eq:qubit2shift}). (b) Time-averaged transition probability of unwanted process $\ket{0 0}  \rightarrow \ket{1 0}$ which is closest to resonance. Parameters:  $\omega_1/(2 \pi) = 12$ GHz, $\omega_2/(2 \pi) = 9$ GHz and $b/(2 \pi) = 150$ MHz.} 
  \label{fig:single_photon_resonance}
\end{figure*}

The incurred local detunings of the qubits due to the modulation cause a shift in the resonance in the transition probabilities such that one needs to detune the frequency of the drive accordingly to $\omega_d = \omega_2 + \delta \omega_2$ to get efficient interactions. In Fig.~\ref{fig:single_photon_resonance1}, this shift in resonance  is shown in the time-averaged transition probability $\ket{00} \rightarrow \ket{01}$, c.f. Eq.~(\ref{eq:time_independent_transition_probability}), as a function of external drive frequency $\omega_d$ and coupling strength $g_c$. In Fig.~\ref{fig:single_photon_resonance2}, the probability of an erroneous process $\ket{00} \rightarrow \ket{10}$ is given for comparison. With the shifted external frequency and in the rotating frame of the qubits with respect to the modified qubit frequencies (i.e. $\omega_i \rightarrow \omega_i + \delta \omega_i$ in Eq.~(\ref{eq:qubit_rot_frame})), the system exhibits a pure interaction $H_{eff} =  J_{zx} \sigma_1^{z} \sigma_2^{x}$.  
Analogously, for $\theta \neq 0$ an effective interaction $H_{eff} = J_{eff} \exp{(i \theta)} \sigma_1^z \sigma_2^+ + \text{H.c.}$ is realized ($J_{eff} \sim J_{zx}$), leading to $H_{eff} = J_{zy} \sigma^z_1 \sigma_2^y$ for $\theta = \pi$.

\subsubsection{Longitudinal modulation at single photon transition} \label{sssec:az_single}
We now show how to engineer a squeezing or hopping interaction in a single-mode driven system with longitudinal modulation, i.e. $a=z$ in Eq.~(\ref{eq:single_mode}), see also Fig.~\ref{fig:circuit}. We first set $\theta=0$. The relevant modulation frequency for squeezing is $\omega_d = \omega_1 + \omega_2$. At this modulation frequency, the two bare Floquet states $\ket{\psi_1 ,  m =  -1 \rangle}$ and $\ket{\psi_4, m=0 \rangle}$ are resonantly coupled through a single photon transition and the relevant manifold for the effective Hamiltonian is $\mathcal{S} = \{\ket{\psi_1 ,  m =  -1 \rangle}$, $\ket{\psi_4 ,  m =  0 \rangle}$, $\ket{\psi_2 , m = -1 \rangle}$, $\ket{\psi_3, m = 0 \rangle}\}$.
These states have quasienergies $\epsilon_1^{(0)} = \epsilon_4^{(0)} = 0$, $\epsilon_2^{(0)} =  -\omega_2$ and $\epsilon_3^{(0)} = \omega_2$. They are, nonetheless, degenerate at zero frequency in the rotating frame defined by $\mathcal{U}_{-} = \exp{( i \omega_2 t/2 (\sigma_1^z - \sigma^z_2) \otimes \mathbb{1})}$, which is identical to the rotating frame of the qubits in the Hilbert space. The Salwen matrix in the subspace $\mathcal{S}$ takes the form,
\begin{equation} \label{eq:structure_squeezing}
 \begin{bmatrix}
  h_{11}(\epsilon)   &0 & 0 & h_{14}(\epsilon)  \\
  0 &h_{22}(\epsilon)   & h_{23}(\epsilon) &0 \\
  0  &h_{32}(\epsilon) &h_{33}(\epsilon) &0  \\
  h_{41}(\epsilon) &0&0&h_{44}(\epsilon) \\
 \end{bmatrix}
\end{equation}
This structure implies that the Hilbert space decouples again into two subspaces: $\mathcal{S}_1 = \{ \ket{\psi_1 \rangle} , \ket{\psi_4 \rangle}\}$ and $\mathcal{S}_2 = \{ \ket{\psi_2 \rangle} , \ket{\psi_3 \rangle}\}$, i.e. as opposed to the transverse modulation $\ket{\psi_1 \rangle}$ is not coupled to $ \ket{\psi_2 \rangle}$ but $ \ket{\psi_4 \rangle}$.
 The effective Hamiltonian for squeezing thus reads,
\begin{equation}
 H_{eff} = \delta \omega_1 \sigma^{z}_1 + \delta \omega_2 \sigma^{z}_2 + J_{s} (\sigma^{+}_1 \sigma^{+}_2 + \text{H.c.}) 
\end{equation}
with the parameters,
\begin{widetext}
\begin{equation}
 \begin{split}
  & \delta \omega_1 =  -\frac{b^2 \eta^2 \omega_1}{\omega_1^2-\omega_2^2} - \frac{b^4 \eta^2 ((2+\eta^2)\omega_1^4 - 4\omega_1^3\omega_2  + (1+3\eta^2) \omega_1^2 \omega_2^2+\omega_2^4)}{\omega_1(\omega_1^2-\omega_2^2)^3} + O(b^6), \\
  & \delta \omega_2 = \frac{- b^2 \eta^2 \omega_2}{\omega_1^2-\omega_2^2} + \frac{b^4 \eta^2 \omega_2 ((2+3 \eta^2) \omega_1^3 - 5 \omega_1^2 \omega_2 + (2+\eta^2) \omega_1 \omega_2^2+\omega_2^3)}{\omega_1(\omega_1^2-\omega_2^2)^3} + O(b^6), \\
  & J_{s} = \frac{2 b^2\eta }{\omega_1+\omega_2} - \frac{ 2 b^4 \eta(2+\eta^2)}{(\omega_1+\omega_2)^3} + O(b^6)
 \end{split}
\end{equation}
\end{widetext}
As an example, for $\omega_1/(2 \pi)= 12$ GHz, $\omega_2/(2 \pi) =  9$ GHz, $b/(2 \pi)=200$ MHz and $g_c/(2 \pi)=500$ MHz ($\eta=2.5$), we get to the fourth order $\delta \omega_1/(2 \pi) =  46.46$ MHz (exact: $46.52$ MHz), $\delta \omega_2/(2 \pi) =  -34.57$ MHz (exact: $-34.63$ MHz) and $J_s/(2 \pi) = 9.51$ MHz (exact: $9.51$ MHz).

Alternatively,  for $\omega_d = \omega_1 - \omega_2$ an effective hopping $J_h \sigma_1^+ \sigma_2^- + \text{H.c.}$ is enabled through a one photon transition. For this working point, the two states $\ket{\psi_2 \rangle}$ and $\ket{\psi_3 \rangle}$ are coupled and the quasienergies of the states in $\mathcal{S}$ after subtracting an irrelevant constant read $\epsilon_2^{(0)} = \epsilon_3^{(0)} = 0$, $\epsilon_1^{(0)} =  \omega_2$ and $\epsilon_4^{(0)} = -\omega_2$.  The relevant states $\mathcal{S}_1 = \{ \ket{\psi_1 ,  m =  -1 \rangle}, \ket{\psi_4 ,  m =  0 \rangle} \}$ and $\mathcal{S}_2 = \{ \ket{\psi_2 , m = -1 \rangle}, \ket{\psi_3, m = 0 \rangle} \}$ are degenerate at zero energy in the rotating frame set by $\mathcal{U}_{+} = \exp{[ i (\omega_2 t/2) (\sigma_1^z  + \sigma^z_2) \otimes \mathbb{1}]}$ which is identical to the rotating frame of the qubits in the Hilbert space.
The effective Hamiltonian for the hopping thus is,
\begin{equation}
  H_{eff} = \delta \omega_1 \sigma^{z}_1 + \delta \omega_2 \sigma^{z}_2 + J_{h} (\sigma^{+}_1 \sigma^{-}_2 + \text{H.c.}) 
\end{equation}
with,
\begin{widetext}
\begin{equation}
 \begin{split}
  & \delta \omega_1 =  \frac{b^2 \eta^2 \omega_1}{\omega_1^2-\omega_2^2} - \frac{b^4 \eta^2 ((2+\eta^2)\omega_1^4 + 4\omega_1^3\omega_2  + (1+3 \eta^2) \omega_1^2 \omega_2^2+\omega_2^4)}{\omega_1(\omega_1^2-\omega_2^2)^3} + O(b^6), \\
  & \delta \omega_2 = \frac{- b^2 \eta^2 \omega_2}{\omega_1^2-\omega_2^2} + \frac{b^4 \eta^2 \omega_2 ((2+3 \eta^2) \omega_1^3 + 5 \omega_1^2 \omega_2 + (2+\eta^2) \omega_1 \omega_2^2 - \omega_2^3)}{\omega_1(\omega_1^2-\omega_2^2)^3} + O(b^6), \\
  & J_{h} = \frac{-2 b^2\eta }{\omega_1-\omega_2} - \frac{ 2 b^4 \eta(2+\eta^2)}{(\omega_1-\omega_2)^3} + O(b^6)
 \end{split}
\end{equation}	
\end{widetext}

For $\omega_1 /(2 \pi)=  12$ GHz, $\omega_2/(2 \pi)= 9$ GHz, $g_c/(2 \pi) = 250$~MHz and $b/(2 \pi)=100$ MHz, we get $\delta \omega_1/(2 \pi) = 11.80$ MHz, $\delta \omega_2/(2 \pi) =  -8.83$ MHz and $J_h/(2 \pi) = -16.51$ MHz. For a  $\theta \neq 0$ one could also incorporate a static gauge field into the squeezing or hopping i.e. $J_s  \rightarrow J_s e^{i \theta}$ and $J_h  \rightarrow J_h e^{i \theta}$.

The degeneracy of low-energy states can however be achieved not only for a single photon transitions but also for multiple-photon exchange with the drive. While this typically leads to lower strength of the effective interactions, it only requires lower frequencies for the driving fields, which can be of a practical advantage. We therefore briefly discuss the engineering of interactions via multi-photon transitions in appendix \ref{ssec:multi-photon}.

\subsection{Bimodal driven system} \label{ssec:bimodal_driven_system}
We now consider a scenario in which both of the linearly coupled qubits are driven, resulting in a bimodal drive with frequencies $\omega_{d1} = \omega_1 + \omega_2$ and $\omega_{d2} = \omega_1 - \omega_2$, see Fig.~\ref{fig:circuit}. The explicit form of the drive Hamiltonian is given in Eq.~(\ref{eq:bimodal_mod}). We will here show that the interactions $\sigma^x\sigma^x$, $\sigma^y\sigma^y$, $\sigma^x\sigma^y$ and $\sigma^z \sigma^z$ can be realized in this system using appropriate forms of the drive. 

Based on the generalization of single-mode Floquet theory to many-modes discussed in Sec. \ref{ssec:many_mode_Floquet}, we can represent this quasiperiodic Hamiltonian in the composite Hilbert space $\mathcal{H}_c = \mathcal{H} \otimes \mathcal{T}_1 \otimes \mathcal{T}_2$ via the Floquet matrix given in Eq.~(\ref{eq:many_mode_Floquet_Hamiltonian}), where the non-vanishing Fourier components read,
\begin{align}
	&\tilde{H}^{(0,0)}  = \overline{H} \\
	&\tilde{H}^{(1,1)} = (\tilde{H}^{(-1,-1)})^* = \sum_{j=1,2} \frac{b_{j1}}{2} \exp(i \theta_{j1}) \sigma_j^a\\
	&\tilde{H}^{(1,-1)}  = (\tilde{H}^{(-1,1)})^*= \sum_{j=1,2} \frac{b_{j2}}{2} \exp(i \theta_{j2}) \sigma_j^a
\end{align}
with $a=x$ for the transverse driving scheme and $a=z$ for the longitudinal driving scheme, c.f. Eq.~(\ref{eq:bimodal_mod}).
  
 In the non-interacting limit, $g_c = b_{ij} = 0$, the states $\ket{\psi_{\alpha},m_1,m_2\rangle}$ are the eigenstates of $H_{F0} \equiv H_{F}\big|_{g_c ,b_{ij} = 0}$ with quasienergies $\epsilon_{(\alpha,m_1,m_2)} = E^{(0)}_{\alpha} + m_1 \omega_1 + m_2 \omega_2$. In particular, the four states $\ket{\psi_1\rangle} \equiv \ket{\psi_1,m_1=-1,m_2=-1\rangle}$,  $\ket{\psi_2 \rangle} \equiv \ket{\psi_2,m_1=-1,m_2=0\rangle}$,  $\ket{\psi_3 \rangle} \equiv \ket{\psi_3,m_1=0,m_2=-1\rangle}$ and  $\ket{\psi_4 \rangle} \equiv \ket{\psi_4,m_1=0,m_2=0\rangle}$ have energy zero and there is a gap of size $\omega_1-\omega_2$ between these degenerate states and other higher energy states. For interactions such that $|g_c|,|b_{ij}| \ll \omega_1-\omega_2$, the effective dynamics can be described in terms of these sates and we can adiabatically eliminate all other Floquet states. For the derivation of the effective Hamiltonian one calculates the scattering matrix $T$ using the Green's function $\mathcal{G}_{\mathcal{P}}$ and perturbation $V = H_F - H_{ F0}$ as in Sec. \ref{ssec:many_mode_Floquet}. We now apply this procedure to the two types of perturbations, transverse and longitudinal bimodal modulation.
 
\subsubsection{Longitudinal driving scheme} \label{sssec:longitudinal_bimodal_driving_schemes}
In the case of a longitudinal drive with $H_d(t)=\sum_{i=1,2} F_i \sigma_i^z$ the matrix equation has the same structure as Eq.~(\ref{eq:structure_squeezing}), where the subspace spanned by $\mathcal{S}_1 = \{ \ket{\psi_1 \rangle},  \ket{\psi_4 \rangle} \}$ decouples from that spanned by $\mathcal{S}_2 = \{ \ket{\psi_2 \rangle},  \ket{\psi_3 \rangle} \}$ and we can solve for quasienergies in each subspace independently. 

We first assume $\theta_1 = \theta_2 = 0$. In order to expand in a single variable, we define $\eta_{ij} = b_{ij}/g_c$. Using the ansatz of Eq.(\ref{eq:quasienergy_expansion}) for the quasienergies and expanding the matrix elements, one can verify that $\kappa_{\alpha,1} = 0$ and $\kappa_{\alpha,2} = \pm b^2/(\omega_1+\omega_2) + O(b^4)$ (for the two states of $\mathcal{S}_1$, $+(-)$ for $\alpha=1(4)$) and $\kappa_{\alpha,2}=\pm b^2/(\omega_1-\omega_2)+O(b^4)$ (for $\mathcal{S}_2$, $+ (-)$ for $\alpha = 2 (3)$). Including $\kappa_{\alpha,2}$, the effective Hamiltonian is exact to fourth order in the amplitudes $b_{ij}$ and the coupling $g_c$. It reads,
\begin{equation}
H_{eff} = \delta \omega_1 \sigma_1^z + \delta \omega_2 \sigma_2^z + J_{xx} \sigma_1^x\sigma_2^x + J_{yy} \sigma_1^y \sigma_2^y
\end{equation}
where,
\begin{equation} \label{eq:param_xx_yy}
 \begin{split}
  & \delta \omega_1 = \frac{b^2 \omega_1}{\omega_1^2-\omega_2^2} + O(b^4), \\
  & \delta \omega_2 = -\frac{b^2 \omega_2}{\omega_1^2-\omega_2^2} + O(b^4), \\
  & J_{xx} = b^2 (\frac{-\eta_{12} + \eta_{22}}{\omega_1-\omega_2} + \frac{\eta_{11}+\eta_{21}}{\omega_1+\omega_2}) + O(b^4), \\
  & J_{yy} = b^2 (\frac{\eta_{12} - \eta_{22}}{\omega_1-\omega_2} + \frac{\eta_{11}+\eta_{21}}{\omega_1+\omega_2})+O(b^4)
 \end{split}
\end{equation}
To leading order, the frequency shifts $\delta \omega_i$ are independent from the external modulations and originate from the capacitive coupling. These shifts can be absorbed in the effective Hamiltonian by detuning the external drives according to $\omega_{d1} \rightarrow \omega_{d1} + \delta \omega_1 + \delta \omega_2$ and $\omega_{d2} \rightarrow \omega_{d2} + \delta \omega_1 - \delta \omega_2$, which sets the external drives back into resonance with the desired processes. The rotating frame of the qubits also needs to be re-defined with respect to the modified frequencies, i.e. $\omega_i \rightarrow \omega_i + \delta \omega_i$.

\begin{table}[t]
 \begin{tabular}{llll}
\hline
 & \multicolumn{3}{r}{Parameters (MHz)} \\
\cline{2-4}
    & $\delta \omega_1/(2 \pi)$ & $\delta \omega_2/(2 \pi)$ & $J_{xx}/(2 \pi)$ \\
\hline
GSM (2nd order)      &  $15.052$     & $-10.662$   & $13.170$     \\
GSM (4th order)       &  $14.942$     & $-10.557$   &$13.129$       \\
Exact result       & $14.944$    & $-10.559$ & $13.129$     \\
\hline
\end{tabular}
\caption{Numerical values for frequency shifts $\delta \omega_i$ ($i=1,2$) and interaction strength $J_{xx}$. Parameters: $\omega_1/(2 \pi)= 12$GHz, $\omega_2/(2 \pi)= 8.5$GHz, $b/(2 \pi)=300$MHz, $\eta_{11}=\eta_{21}=0.75$, $\eta_{12}=0.256$, $\eta_{22}=0$. GSM means Generalized Slawen method. \label{tab:Jxx}}
\end{table}

By choosing the values for $\eta_{ij}$ such that $(\eta_{12} - \eta_{22})/(\omega_1-\omega_2) = - (\eta_{11}+\eta_{21})/ (\omega_1+\omega_2)$, we get $J_{yy} \approx 0$ and the interaction is purely $\sigma_1^x\sigma_2^x$. In turn, if  $(\eta_{12} - \eta_{22})/(\omega_1-\omega_2) =  (\eta_{11}+\eta_{21})/(\omega_1+\omega_2)$, then $J_{xx} \approx 0$ and a pure  $\sigma_1^y\sigma_2^y$ interaction is implemented (see table \ref{tab:Jxx} for a numerical example). Finally, for $\theta_{ji} = \pi/2$, the effective Hamiltonian takes the form,
\begin{equation}
 H_{eff} = \delta \omega_1 \sigma_1^z + \delta \omega_2 \sigma_2^z + J_{xy} \sigma_1^x \sigma_2^y + J_{yx} \sigma_1^y\sigma_2^x 
\end{equation}
and a mixed interaction is implemented. The parameters of this Hamiltonian can be obtained from Eq.~(\ref{eq:param_xx_yy}) by replacing $J_{xx} \rightarrow J_{xy}$ and $J_{yy} \rightarrow J_{yx}$.
\subsubsection{Transverse driving scheme}
In the case of a transverse driving scheme with $H_d(t)=\sum_{i=1,2} F_i \sigma_i^x$, the off-diagonal elements of the effective matrix are zero, and therefore the effective Hamiltonian is written as,
\begin{equation}
 H_{eff} = \delta \omega_1 \sigma_1^z + \delta \omega_2 \sigma_2^z + J_{zz} \sigma_1^z \sigma_2^z
\end{equation}
To third order in interaction strength $b$ the parameters read,
\begin{widetext}
\begin{equation}
 \begin{split}
  & \delta \omega_1 = \frac{-b^2(\eta_{11}^2-\eta_{12}^2)}{\omega_2}+\frac{b^2 \omega_1}{\omega_1^2-\omega_2^2} + \frac{b^2(\eta_{11}^2(2\omega_1-\omega_2)+\eta_{12}^2(2\omega_1+\omega_2)}{4\omega_1^2-\omega_2^2} + O(b^4), \\
  & \delta \omega_2 =  \frac{-b^2(\eta_{21}^2-\eta_{22}^2)}{\omega_1}-\frac{b^2 \omega_2}{\omega_1^2-\omega_2^2} + \frac{b^2(\eta_{21}^2(\omega_1-2\omega_2)-\eta_{22}^2(\omega_1+2\omega_2)}{\omega_1^2-4\omega_2^2} + O(b^4), \\
  & J_{zz} =  \frac{4 b^3(\eta_{11}\eta_{21}-\eta_{12}\eta_{22})}{\omega_1\omega_2}+\frac{4b^3\eta_{12}\eta_{22}}{3\omega_1(2\omega_1-\omega_2)} + \frac{4b^3\eta_{11}\eta_{21}}{3\omega_1(2\omega_1+\omega_2)} - \frac{8 b^3 \eta_{12} \eta_{22}}{3\omega_1(\omega_1-2\omega_2)}  - \frac{8 b^3 \eta_{11} \eta_{21}}{3\omega_1(\omega_1+2\omega_2)} + O(b^4)
 \end{split}
\end{equation}
\end{widetext}
For $\omega_1/(2\pi) = 9$GHz, $\omega_2/(2\pi) = 5$GHz, $b/(2 \pi) = 400$MHz, $\eta_1=\eta_3 = 0.6$ and $\eta_2 = \eta_4 = 0$ we here for example get $J_{zz}/(2\pi) = 1.8$MHz.
 
\section{Driven coupling scheme} \label{sec:driven_coupling}
In Sec. \ref{sec:driven_qubit_schemes}, we considered a linear coupling of the form $g_c (a_1- a_1^{\dagger})(a_2 -a_2^{\dagger})$ between two Transmons and discussed the possibility of generating spin-spin interaction by using an appropriate single or bimodal driving of the qubits. The implemented interactions are at least second-order in modulation strength and/or coupling strength such that interactions of the order of 10-15MHz are realized with that setting. 

\begin{figure}
  \centering
  \includegraphics[width=0.4\textwidth]{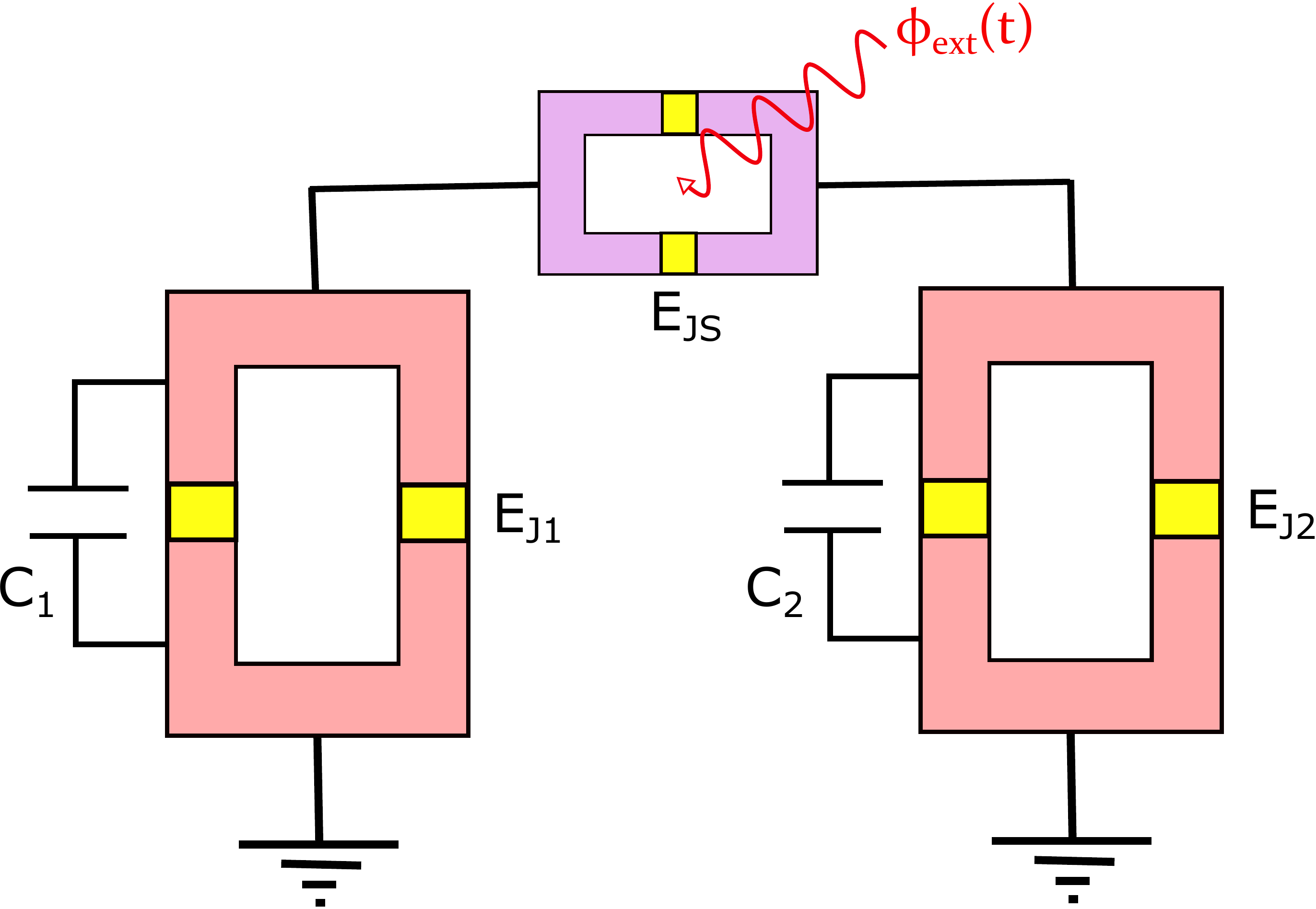}
  \caption{Two Transmons with Josephson energy $E_{Jj}$ and capacitance $C_j$ coupled via driven SQUID with Josephson energy $E_{Js}$ and negligible capacitance. The form of the bimodal external flux $\phi_{ext}(t)$ is defined in Eq.~(\ref{eq:external_flux}).
  \label{fig:bimodal_nl}}
\end{figure}

In this section, we turn to a different type of coupling, namely a driven nonlinear coupling, see Fig.~\ref{fig:bimodal_nl}. As we will show in the following, higher effective interaction strengths can be achieved with this nonlinear coupling as the effective interaction is first-order in the modulation strength. In contrast to the linear coupling, parity-breaking interactions can however not be realized in this setup since all of the contributing terms in the Hamiltonian are parity-conserving. This coupling circuit also requires some additional control circuitry, which can make the driven qubit circuit a better choice for applications that require only moderate interaction strength.

We here consider two Transmon qubits with Josephson energies $E_{Ji}$ and capacitances $C_i$ coupled via a dc-SQUID with Josephson energy $E_{Js}$ and capacitance $C_s$, see Fig.~\ref{fig:bimodal_nl}. We further assume that the coupling SQUID is modulated via an external time-dependent flux $\phi_{ext}(t)$. The Hamiltonian of the circuit reads,
\begin{align}
	H & = \sum_{j=1,2} \left(E_{Jqj}  \cos(\phi_j) + E_{Cqj} n_j^2 \right) \\
	& + E_{cc} n_1 n_2 + E_{Js} \cos(\phi_{ext}/2) \cos(\phi_1 - \phi_2) \nn
\end{align}
For the external flux, we assume a superposition of a constant and a bimodal contribution,
\begin{equation} \label{eq:external_flux}
\begin{split}
 &\phi_{ext}(t) = \phi_{dc} +  F(t), \\ 
 & F(t) = 2\phi_{ac,1} \cos(\omega_{d1}t+ \theta) + 2\phi_{ac,2} \cos(\omega_{d2}t )
\end{split}
 \end{equation}
where $\omega_{d1} = \omega_1 +\omega_2$, $\omega_{d2} = \omega_1 - \omega_2$ and for $\phi_{ac} \ll \phi_{dc}$, we have $\cos(\phi_{ext}/2) \approx \cos(\phi_{dc}/2) + \sin(\phi_{dc}/2) F(t)$.
In the single excitation subspace, the quantized Hamiltonian reads,
\begin{equation} \label{eq:bimodal_nl_Hamiltonian}
 \begin{split}
 H &= H_0 + H_1(t) \\
 H_0 &= \sum_{j=1,2}  \omega_j \sigma_j^+ \sigma_j^- \\
 H_1(t) &=  \left[ \cos\left(\frac{\phi_{dc}}{2}\right) + \sin\left(\frac{\phi_{dc}}{2}\right)F(t) \right] H_1'
 \end{split}
\end{equation}
with,
\begin{equation}
	 H_1' =  \sum_{j=1,2} g_j \sigma_j^+ \sigma_j^-  + g_{x}  \sigma_1^x  \sigma_2^x + g_{z} \sigma^z_1 \sigma^z_2,
\end{equation}
where $g_j = E_{Js} \overline{\phi}_j^2/2$, $g_x = -E_{Js} \overline{\phi}_1 \overline{\phi}_2$, $g_z = E_{Js} \overline{\phi}_1^2 \overline{\phi}_2^2/4$ and $g_c = -E_{cc}/(4 \overline{\phi}_1 \overline{\phi}_2)$.
We have here assume a weak capacitive coupling, $g_c \ll g_x, g_z, g_j$, and neglected a term  $g_{c} \sigma_1^y \sigma_2^y$ in the Hamiltonian $H_1$. 

To discuss the generation of effective spin-spin interactions in this circuit, we first outline the idea using a more heuristic argument based on a Rotating Wave Approximation and then turn to derive more precise expressions with our Floquet engineering approach. Before entering the discussion of the derivation, we summarize the results for this section.

\subsection{Summary of results}
For a single and bimodal drives, applied to the coupling circuit, the following interactions may be generated:
\begin{enumerate}
	\item For $\phi_{dc} \neq n \pi$ with odd integer $n$, an interaction $J_{zz} \sigma_1^z \sigma_2^z$ with $J_{zz} \sim g_z$ is generated.
	\item If the coupler is driven with $\omega_{d,1} = \omega_1 - \omega_2$, the hopping term $J_{+-} \phi_{ac,1} \sigma_1^+ \sigma_2^- + \text{H.c.}$ with $J_{+-} \sim g_x$ is enabled.
	\item If the coupler is driven with $\omega_{d,2} = \omega_1 + \omega_2$, the squeezing terms $J_{++} \phi_{ac,2} \sigma_1^+ \sigma_2^+ + \text{H.c.}$ with $J_{++} \sim g_x$are enabled.
	\item If $\phi_{ac} = \phi_{ac,1} = \phi_{ac,2}$, the interaction $J_{xx} \phi_{ac} \sigma_1^x \sigma^x_2$ is generated for $\theta=0$ and the interaction $J_{yy} \phi_{ac} \sigma_1^y \sigma^y_2$ for $\theta=\pi$, where $J_{xx} \sim J_{yy} \sim g_x$. 
\end{enumerate}
The precise values for these interactions are given in Eq.~(\ref{eq:drive-couple-floquet-para}).

\subsection{Rotating Wave Approximation (RWA)}
Let us first assume that the oscillating drive is turned off, $\phi_{ac,1}=\phi_{ac,2} = 0$. In this case, the Hamiltonian reads $H = \sum_{j=1,2}  \tilde{\omega}_j \sigma_j^+ \sigma_j^- +   \tilde{g}_{x}  \sigma_1^x  \sigma_2^x + \tilde{g}_{z} \sigma^z_1 \sigma^z_2$, where $\tilde{\omega}_j = \omega_j + \cos(\phi_{dc}/2) g_j$ is the modified qubit transition frequency, $\tilde{g}_x = \cos(\phi_{dc}/2) g_x$ and $\tilde{g}_z = \cos(\phi_{dc}/2) g_z$.
In the rotating frame of the qubits defined by $U_{qj}=\exp(-i \tilde{\omega}_j t \sigma^+_j \sigma^-_j)$, the Hamiltonian and the spin operators transform as $H \rightarrow H-\sum_{j=1,2}  \tilde{\omega}_j \sigma_j^+ \sigma_j^-$ and $\sigma^{\pm}_j \rightarrow \exp(\pm i \tilde{\omega}_j t) \sigma^{\pm}_j$. Therefore, the term $\tilde{g}_x \sigma^x_1 \sigma^x_2$ rotates at the two frequencies $\tilde{\omega}_{d1} = \tilde{\omega}_1 + \tilde{\omega}_2$ and $\tilde{\omega}_{d2} = |\tilde{\omega}_1  - \tilde{\omega}_2|$. Provided  $\tilde{g}_x \ll \tilde{\omega}_{d1}, \tilde{\omega}_{d2}$, which is typically the case, we can therefore discard it. The Hamiltonian is then well-approximated by the non-rotating term $H \approx \tilde{g}_z \sigma^z_1 \sigma^z_2$. 

If instead, we apply the oscillating flux described in Eq.~(\ref{eq:external_flux}) and set $\phi_{dc} = \pi$, the term $\sigma^+_j \sigma^-_j$ and $\sigma^z_1 \sigma^z_2$ in $H_1'$ rotate at the frequencies $\omega_{d1}$ and $\omega_{d2}$ in the rotating frame of the qubits. For $g_j,g_z \ll \omega_{d1}, \omega_{d2}$ these terms can then be neglected in a rotating wave approximation. The only non-rotating contribution is then given by the term $F(t)g_x\sigma^x_1 \sigma^x_2$, which, in the rotating frame, reads
\begin{equation}\label{eq:eff_H_coup_drive_1}
 g_x e^{i \theta}\phi_{ac,1} \sigma^+_1 \sigma^+_2 + g_x\phi_{ac,2}\sigma^+_1 \sigma^-_2 + \text{H.c.}
\end{equation}
where we have dropped rotating terms which oscillate at $2 \omega_{dj}$ and $2\omega_j$ for $j=1,2$, since $g_x \ll 2 \omega_{dj}, 2\omega_j$. 

Therefore, if $\phi_{ac} = \phi_{ac,1} = \phi_{ac,2}$, we get $H \approx g_x \phi_{ac} \sigma_1^x \sigma^x_2$ for $\theta=0$ and $H \approx g_x \phi_{ac} \sigma_1^y \sigma^y_2$ for $\theta=\pi$. We also note that for the two single mode driving cases, $\phi_{ac,1}=0$ or $\phi_{ac,2}=0$ a hopping or a squeezing term would be enabled, see also \cite{Collodo18}.

\subsection{Bimodal Floquet Theory}
We now derive more accurate expressions for the effective interactions generated by the drive described in Eq.~(\ref{eq:external_flux}) with $\phi_{dc} = \pi$ via our Floquet engineering approach.
The infinite-dimensional matrix $H_F$ to represent this bimodal driven system in Floquet space is given in Eq.~(\ref{eq:many_mode_Floquet_Hamiltonian}) where here, 
\begin{align}
	&\tilde{H}^{(0,0)}  = H_0  \\
	&\tilde{H}^{(1,1)} =(\tilde{H}^{(-1,-1)})^* = \phi_{ac,1}e^{i \theta} H_1'\\
	&\tilde{H}^{(1,-1)}  =(\tilde{H}^{(-1,1)})^* =  \phi_{ac,2} H_1'
\end{align}
In the limit $|g_j|,|g_x|,|g_z| \ll |\omega_1-\omega_2|$, we can derive an effective Hamiltonian in terms of the four states $\ket{\psi_1\rangle} \equiv \ket{\psi_1,m_1=-1,m_2=-1\rangle}$,  $\ket{\psi_2 \rangle} \equiv \ket{\psi_2,m_1=-1,m_2=0\rangle}$,  $\ket{\psi_3 \rangle} \equiv \ket{\psi_3,m_1=0,m_2=-1\rangle}$ and  $\ket{\psi_4 \rangle} \equiv \ket{\psi_4,m_1=0,m_2=0\rangle}$. Using Salwen's method, the effective matrix again decouples into two subspaces $S_1=\{ \ket{\psi_1 \rangle},   \ket{\psi_4 \rangle} \}$ and $S_2=\{ \ket{\psi_2 \rangle},   \ket{\psi_3 \rangle} \}$. For $\phi_{ac} = \phi_{ac,1} = \phi_{ac,2}$ and $\theta=0$, we get the following quasienergies for these subspaces,
\begin{widetext}
\begin{equation}
	\begin{split}
 &\epsilon_{\alpha} = b - \bigg( \frac{1}{\omega_1^2} + \frac{1}{\omega_2^2} +\frac{32\eta_{\alpha}^2}{(\omega_1-\omega_2)^2} +\frac{1+16 \eta_{\alpha}^2}{(\omega_1+\omega_2)^2} \bigg)b^3+O(b^5) \quad \text{for} \quad \alpha=1,2\\
&\epsilon_{\alpha} = -b - \bigg( \frac{3}{\omega_1^2} + \frac{3}{\omega_2^2} +\frac{32\eta_{\alpha}^2}{(\omega_1-\omega_2)^2} +\frac{3+16 \eta_{\alpha}^2 + 32 \eta_z^2}{(\omega_1+\omega_2)^2}\bigg)b^3+O(b^5) \quad \text{for} \quad \alpha=3,4.
 \end{split}
 \end{equation}
 where $\eta_1=\eta_4 = (g_1-g_2)/g_x$, $\eta_2 = \eta_3 =(g_1+g_2)/g_x$, $\eta_z=g_z/g_x$ and $b = g_x \phi_{ac}$. The effective Hamiltonian thus reads,
 \begin{equation}\label{eq:eff_H_coup_drive}
  H_{eff} =\delta \omega_1 \sigma_1^z+ \delta \omega_2 \sigma^z_2 + J_{xx} \sigma_1^x \sigma_2^x + J_{yy} \sigma_1^y \sigma_2^y
 \end{equation}
and the parameters of this Hamiltonian to third order in $b$ are,
\begin{equation} \label{eq:drive-couple-floquet-para}
 \begin{split}
  & \delta \omega_1 = \frac{b^2}{4}\bigg( \frac{2}{\omega_1} + \frac{1}{\omega_1-\omega_2} + \frac{1}{\omega_1+\omega_2} \bigg) -2b^3 (\eta_1-\eta_2) \eta_z \bigg( \frac{1}{\omega_1-\omega_2} + \frac{1}{\omega_1+\omega_2} \bigg), \\
  & \delta \omega_2 = \frac{b^2}{4}\bigg( \frac{2}{\omega_2} - \frac{1}{\omega_1-\omega_2} + \frac{1}{\omega_1+\omega_2} \bigg) -2b^3 (\eta_1+\eta_2) \eta_z \bigg( \frac{1}{\omega_1-\omega_2} + \frac{1}{\omega_1+\omega_2} \bigg), \\
  & J_{xx} = b + b^3 \bigg( \frac{2\eta_z^2-\eta_1^2}{(\omega_1-\omega_2)^2} + \frac{2\eta_z - \eta_2^2}{(\omega_1+\omega_2)^2} \bigg),  \\
  & J_{yy} = b^3 \bigg( \frac{ \eta_1^2}{(\omega_1-\omega_2)^2}  + \frac{1}{2\omega_1\omega_2} - \frac{ \eta_2^2}{(\omega_1-\omega_2)^2}  \bigg)
 \end{split}
\end{equation}
\end{widetext}
The effective Hamiltonian (\ref{eq:eff_H_coup_drive}) differs from Eq.~(\ref{eq:eff_H_coup_drive_1}), which was obtained from a rotating wave approximation, in two aspects: it contains frequency shifts on the qubits and interaction terms of third order. Higher order interaction terms can also be derived if desired. For $\theta = 0$, we can eliminate the $\sigma_1^y\sigma_2^y$ interaction by assuming a slight difference in the driving amplitudes, i.e. by choosing $\phi_{ac,1} = \phi_{ac}$ and $\phi_{ac,2} = \phi_{ac} + \delta \phi_{ac}$. To linear order in $\delta \phi_{ac}$, the strength of the  $\sigma_1^y\sigma_2^y$ interaction is then given by,
\begin{align}
	 J_{yy} & = \left.J_{yy}\right|_{\delta \phi_{ac} = 0} + D_y \delta \phi_{ac} = 0, \quad \text{where}\\
 D_y & = \frac{g_x}{2} \bigg( 1 + \frac{3b^2}{2\omega_1\omega_2} + \frac{6 b^2 \eta_z^2}{(\omega_1-\omega_2)^2} + \frac{6b^2(\eta_2^z-\eta_2^2)}{(\omega_1+\omega_2)^2} \bigg) \nn \\
 & \approx g_x/2 \nn
\end{align}
and $\left.J_{yy}\right|_{\delta \phi_{ac} = 0}$ is given in Eq.~(\ref{eq:drive-couple-floquet-para}). Hence for an amplitude mismatch of $\delta \phi_{ac} \approx 2 J_{yy}/g_{x}$, the $YY$-interaction is suppressed, $J_{yy} = 0$.

\section{Quantum simulation of Kitaev Honeycomb Lattice model} \label{sec:Honeycomb}
\begin{figure*}
\hspace{0.125 \linewidth}
\begin{tikzpicture}
  \definecolor{zz}{RGB}{232,178,240}
  \definecolor{qubit}{RGB}{255,170,170}
  \definecolor{yy}{RGB}{42,212,255}

  \foreach \i in {0,...,5} 
  \foreach \j in {1,...,2} {
  \foreach \a in {90} \draw[zz,line width=0.5mm] (2*\i*cos{30},2*\j+2*\j*sin{30}) -- +(\a:1);}
    
  \foreach \i in {0,...,4} 
  \foreach \j in {1} {
  \foreach \a in {90} \draw[zz,line width=0.5mm] (cos{30} + 2*\i*cos{30}, 1+ 2*\j+sin{30}+2*\j*sin{30}) -- +(\a:1);}
  
  \foreach \i in {0,...,4} 
  \foreach \j in {1,...,2} {
  \foreach \a in {-30} \draw[green,line width=0.5mm] (2*\i*cos{30},2*\j+2*\j*sin{30}) -- +(\a:1);}
  
   \foreach \i in {0,...,4} 
  \foreach \j in {1,...,2} {
  \foreach \a in {-30} \draw[green,line width=0.5mm] (cos{30} + 2*\i*cos{30}, 1+ 2*\j+sin{30}+2*\j*sin{30}) -- +(\a:1);}
  
  \foreach \i in {1,...,5} 
  \foreach \j in {1,...,2} {
  \foreach \a in {-150} \draw[yy,line width=0.5mm] (2*\i*cos{30},2*\j+2*\j*sin{30}) -- +(\a:1);}
    
  \foreach \i in {0,...,4} 
  \foreach \j in {1,...,2} {
  \foreach \a in {-150} \draw[yy,line width=0.5mm] (cos{30} + 2*\i*cos{30}, 1+ 2*\j+sin{30}+2*\j*sin{30}) -- +(\a:1);}
  
  \foreach \i in {0,...,5} 
  \foreach \j in {1,...,2} {
  \fill[qubit] (2*\i*cos{30},2*\j+2*\j*sin{30}) circle (4pt) ;}
    
  \foreach \i in {0,...,4} 
  \foreach \j in {1,...,2} {
  \fill[qubit] (cos{30} + 2*\i*cos{30}, 1+ 2*\j+sin{30}+2*\j*sin{30}) circle (4pt);}
  
  \foreach \i in {0,...,5} 
  \foreach \j in {1,...,2} {
  \fill[orange] (2*\i*cos{30},1+2*\j+2*\j*sin{30}) circle (4pt) ;}
    
  \foreach \i in {0,...,4} 
  \foreach \j in {0,...,1} {
  \fill[orange] (cos{30} + 2*\i*cos{30}, 2+ 2*\j+sin{30}+2*\j*sin{30}) circle (4pt);
  }
  
  \foreach \a in {90} \draw[zz,line width=0.5mm] (11, 3+3*sin{30}) -- +(\a:1);
  \foreach \a in {-30} \draw[green,line width=0.5mm] (11,3+3*sin{30}) -- +(\a:1);
  \foreach \a in {-150} \draw[yy,line width=0.5mm] (11, 3+ 3*sin{30}) -- +(\a:1);
  \fill[qubit] (11, 3+ 3*sin{30}) circle (4pt);
  \fill[orange] (11, 4+ 3*sin{30}) circle (4pt);
  \fill[orange] (11+1*cos{30}, 3+ 2*sin{30}) circle (4pt);
  \fill[orange] (11-1*cos{30}, 3+ 2*sin{30}) circle (4pt);
  \node[text width=3cm] at (12.1,3.4+3*sin{30}) {zz};
  \node[text width=3cm] at (12, 3.1+2*sin{30}) {xx};
  \node[text width=3cm] at (12.9, 3.4+2*sin{30}) {yy};
\end{tikzpicture}
\caption{The Kitaev honeycomb lattice model. Spins sitting on the vertices of a honeycomd lattice constitute the degrees of freedom of the model. Each spin interacts via 3 different color coded interactions $XX$, $YY$ and $ZZ$ with nearest neighbors.}
\label{fig:honeycomb_model}
\end{figure*}
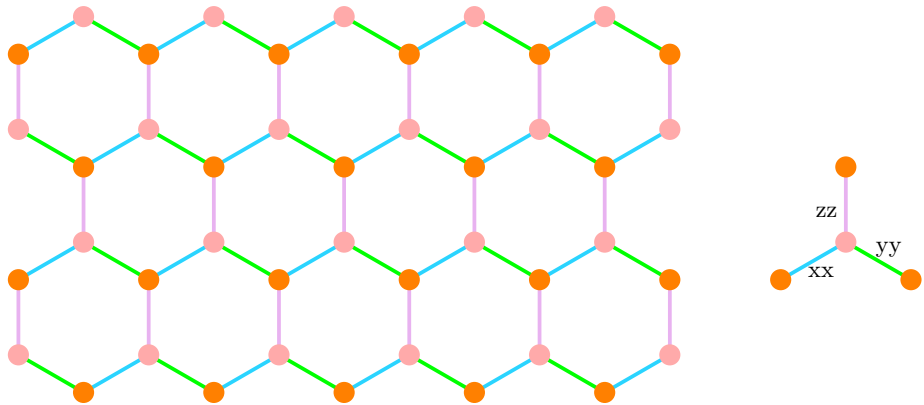

We now put the engineered spin-spin interactions together to develop a proposal for a superconducting quantum simulator for the Kitaev honeycomb lattice model~\cite{Kitaev03,Kitaev09}. In this model, the degrees of freedom are spins at the vertices of a honeycomb lattice, see Fig.~\ref{fig:honeycomb_model}. Each spin is coupled to its three nearest neighbors through three different types of interactions $\sigma^x \sigma^x$, $\sigma^y \sigma^y$ and $\sigma^z \sigma^z$. These interactions are shown as color coded links in Fig.~\ref{fig:honeycomb_model}, where each spin is represented as a dot. The Hamiltonian of the model reads,
\begin{widetext}
\begin{equation}
 H = J_{xx}\sum_{\text{blue links}} \sigma_i^x \sigma_j^x + J_{yy} \sum_{\text{green links}} \sigma_i^y \sigma_j^y + J_{zz}\sum_{\text{pink links}}  \sigma^z_i \sigma^z_j
\end{equation}	
\end{widetext}
In our implementation of this model in superconducting circuits, Transmon qubits form the spin degrees of freedom and we use the two driving schemes discussed in sections \ref{sec:driven_qubit_schemes} and \ref{sec:driven_coupling} to realize the spin-spin interactions, c.f. Fig.~\ref{fig:honeycomb_implementation_linear} and \ref{fig:honeycomb_implementation_nonlinear}. 

For the qubits, we use two sets of transition frequencies, indicated by the pink and orange colors of the dots representing qubits in Fig.~\ref{fig:honeycomb_model}. This ensures that, for each qubit, all three nearest neighbors have a different transition frequency than the qubit itself. Each qubit is thus off-resonant to its three nearest neighbors.

We now discuss the two implementations individually.

\subsection{Implementation based on driven qubit scheme}
In an implementation based on driven qubit scheme, the $XX$ and $YY$ couplings are realized through fixed capacitive couplings as discussed in Sec.~\ref{ssec:bimodal_driven_system}. The $ZZ$ coupling is realized using an unmodulated dc-SQUID which can be tuned via the dc flux through its loop to adjust the coupling strength, see Sec. \ref{sec:driven_coupling}. As we discussed in Sec.~\ref{ssec:bimodal_driven_system}, the $XX$ and $YY$ interactions are second-order while the $ZZ$ coupling is fourth-order in the coupling strength. To ensure that any first-order couplings are ineffective, we therefore assume that nearest neighbor qubits are mutually detuned by several GHz. This large detuning is indicated by the two different colors used for the dots representing qubits in Fig.~\ref{fig:honeycomb_model}. To furthermore suppress next-nearest neighbor interactions, we moreover slightly detune next-nearest neighbor qubits, i.e. any two qubits of the same color in Fig.~\ref{fig:honeycomb_model} which share a common neighbor.

In this scheme based on qubit modulations, each qubit $i$ of the lattice is modulated via a longitudinal drive $F_i(t)\sigma_i^z$. To determine the form of $F_i(t)$, let us first consider two neighboring qubits $i$ and $j$ that form a $XX$ or $YY$ link. Following the discussion in Sec.~\ref{ssec:bimodal_driven_system}, engineering these interactions requires both frequencies $\omega_i \pm \omega_j$. For a pair of qubits, a $XX$ or $YY$ interaction can however be generated by driving one qubit at the sum and the other at the difference of the two transition frequencies. We can therefore choose,
\begin{equation} \label{eq:driven-qubit-freq-choice-1}
	F_i = b_{ij} \cos[(\omega_i + \omega_j) t], \quad 
	F_j = b_{ji} \cos[(\omega_i - \omega_j) t]
\end{equation}
where $b_{ij}$ is the amplitude of the drive applied to qubit $i$ to engineer the interaction with qubit $j$ and vice versa for $b_{ji}$. The drive amplitudes need to fulfill,
\begin{equation} \label{eq:driven-qubit-amp-choice}
	\frac{b_{ij}}{\omega_i + \omega_j} = \pm \frac{b_{ji}}{\omega_i - \omega_j}
\end{equation}
with the $+(-)$ sign to generate a pure $XX(YY)$ interaction.  

Since, in the honeycomb model, each spin has a $XX$ interaction with one neighbor and a $YY$ interaction with another neighbor, and because we assume that these neighbors have different transition frequencies, four different modulation frequencies are required to generate both interactions. For qubit $i$, that forms $XX$ or $YY$ interactions with qubits $j$ and $j'$, these frequency components read,
\begin{equation}
	\omega_i \pm \omega_j \quad \text{and} \quad  \omega_i \pm \omega_{j'}. 
\end{equation}
Following the idea presented in Eq.~(\ref{eq:driven-qubit-freq-choice-1}) for one qubit pair, we can however work with a reduced number of frequency components applied to a qubit and use the drive,
\begin{equation} \label{eq:driven-qubit-freq-choice-2}
 F_i(t) = b_{ij} \cos[(\omega_i + \omega_j) t] + b_{ij'} \cos[(\omega_i - \omega_{j'}) t]
\end{equation}
for qubit $i$. Here the first component of the drive is responsible for generating the interaction with qubit $j$ and the second component for the interaction with qubit $j'$. Note that, due to the assumed detunings between next-nearest neighbors $j$ and $j'$, the first (second) component does not affect the interaction between $i$ and $j'$ ($j$). 

As the drive in Eq.~(\ref{eq:driven-qubit-freq-choice-2}) is not symmetric with respect to the neighbors of the qubit $i$, the positions of the neighbors $j$ and $j'$ need to be fixed consistently for all qubits. We thus choose qubit $j (j')$ to be to the right (left) of qubit $i$. One could equally choose the opposite convention, but it is important to stick to a uniform convention when applying the drive  (\ref{eq:driven-qubit-freq-choice-2}) to qubits. For example, in Fig.~\ref{fig:honeycomb_implementation_linear}, qubit $i$ is to the left of qubit $j$, so the drive $F_j$, applied to qubit $j$ has a frequency component $\omega_j - \omega_i$. On the other hand, qubit $i$ is to the right of qubit $j'$, so that the drive $F_{j'}$ applied to qubit $j'$ has a frequency component $\omega_{j'} + \omega_i$. In this way, each pair of qubits is driven by both the difference and sum of their transition frequencies and $XX$ or $YY$ interactions are implemented.

\subsubsection{Numerical test of the approximations}
\begin{figure}
  \centering
  \subfloat[]{\includegraphics[width=0.46\textwidth]{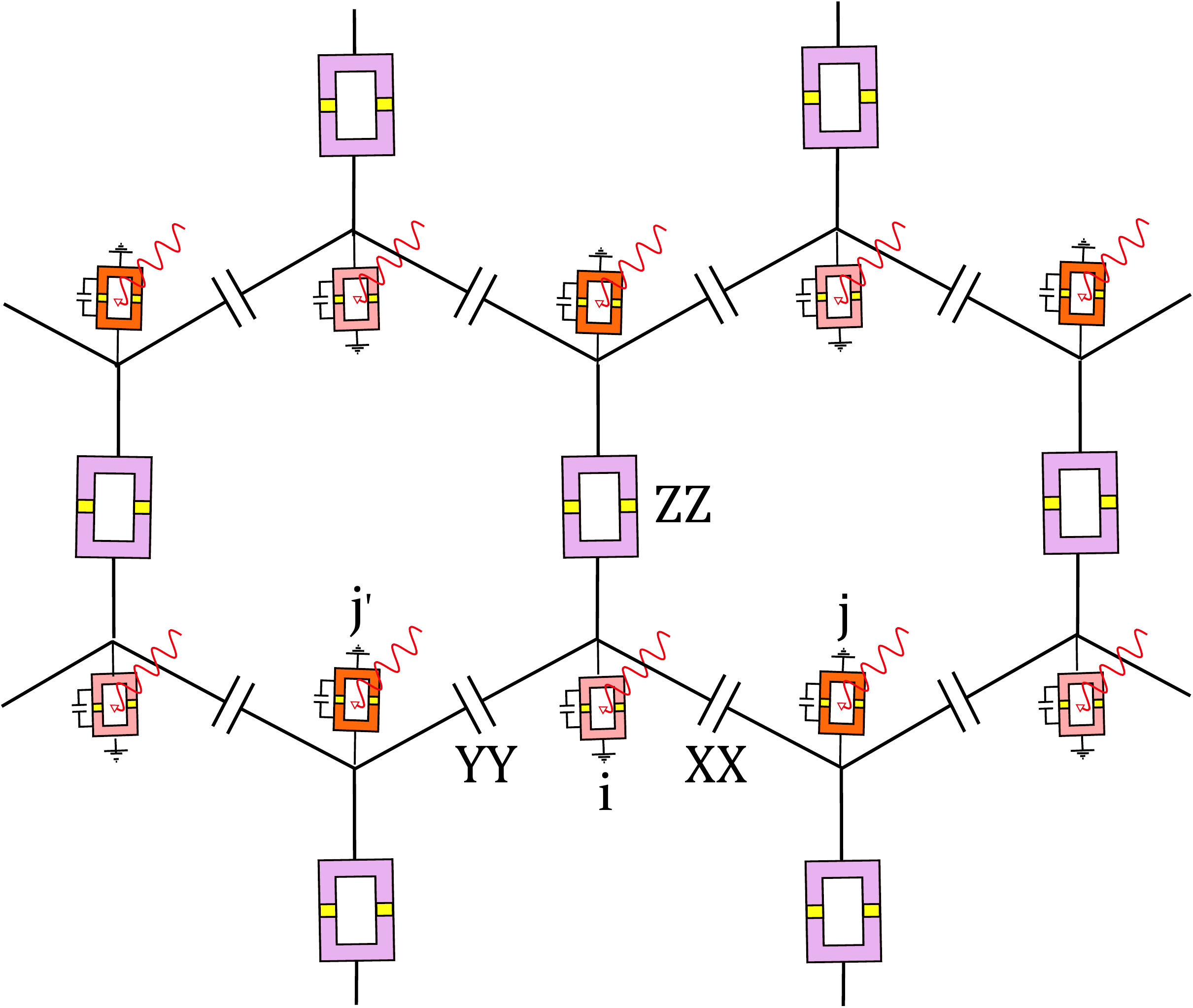}\label{fig:f3}}\\
  \subfloat[]{\includegraphics[width=0.46\textwidth]{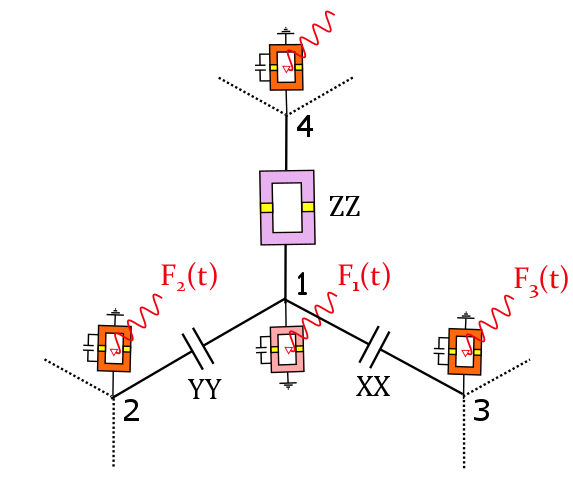}\label{fig:f4}}
  \caption{ (a) Implementation of a honeycomb cell in quantum simulation of the Kitaev model on a honeycomb lattice using driven qubit scheme (b) A typical qubit and three adjacent neighbors used for numerical simulations.}
  \label{fig:honeycomb_implementation_linear}
\end{figure}

To numerically investigate the accuracy of the approximations we used, we now consider a four-qubit module of the honeycomb lattice which consists of a central qubit and its three nearest neighbors, see Fig.~\ref{fig:f4}.

 Via numerical simulations we verify, for this setup, (i) that the drive in Eq.~(\ref{eq:driven-qubit-freq-choice-2}) does generate the required interactions for the honeycomb lattice and (ii) that this drive does not generate any long-range interactions that are not present in the honeycomb model.
 
For the considered four-qubit module, we consider the drives,
\begin{equation} \label{eq:drives}
\begin{split}
 & F_1(t) = b_{12} \cos(\omega_1 - \omega_2) + b_{13} \cos(\omega_1 + \omega_3), \\
 & F_2(t) = b_{21} \cos(\omega_1 + \omega_2), \\
 & F_3(t) = b_{31} \cos(\omega_1  - \omega_3)
 \end{split}
\end{equation}
where we have neglected the drive acting on qubit 4, c.f. Fig.~\ref{fig:f4}, since the $XX$ and $YY$ interactions of this qubit are not contained in the considered subsystem. For the same reasons we also neglected the other components of $F_2$ and $F_3$, because qubit 2 only has a $YY$ interaction and qubit 3 only a $XX$ interaction in the considered subsystem. Due to the chosen pattern of transition frequencies all omitted driving components are off resonant for all couplings between the considered qubits and can therefore safely be neglected.

According to the relation (\ref{eq:driven-qubit-amp-choice}), we choose the drive amplitudes to fulfill,
\begin{equation}
	\frac{b_{12}}{\omega_1 - \omega_2} = \frac{b_{21}}{\omega_1+\omega_2} \quad \text{and} \quad \frac{b_{13}}{\omega_1 + \omega_3} = -\frac{b_{31}}{\omega_1 - \omega_3}
\end{equation}
As an example for the parameters of the minimal circuit, we choose $\omega_1/(2 \pi) =  6.1$ GHz, $\omega_2/(2 \pi) = 9.6$ GHz, $\omega_3/(2 \pi)= 9.1$ GHz and $\omega_4/(2 \pi)= 9.9$ GHz for the transition frequencies of the four qubits, $g_c/(2 \pi) = 200$ MHz for the capacitive couplings [c.f. Eq.~(\ref{eq:hamiltonian_capacitive_coupling})], $g_x/(2 \pi)=200$ MHz and $g_z/(2 \pi) = 10 $ MHz for the $ZZ$ coupling of the SQUID, c.f. Eq.~(\ref{eq:bimodal_nl_Hamiltonian}) for $F(t) \equiv 0$. For the coefficients $b_{ij}$ in Eq.~(\ref{eq:drives}), we assume $b_{12}/(2 \pi) = 213.3$ MHz, $b_{21}/(2 \pi) = -48$ MHz, $b_{13}/(2 \pi) = -41.7$ MHz and $b_{31}/(2 \pi) = 204.1$ MHz. The Floquet matrix of this minimal circuit has four individual modes with four qubit degrees of freedom. We truncate the infinite space of the drive to 9 states ($n=-4$ to $n=4$) for the numerical calculation of the effective Hamiltonian. This results in the Hamiltonian,
\begin{equation} \label{eq:effham_4qubit}
 H_{eff} = \sum_{j=1}^4 \delta \omega_j \sigma_j^z + J_{xx} \sigma^x_1 \sigma^x_3 + J_{yy} \sigma^y_1 \sigma^y_2 + J_{zz} \sigma^z_1 \sigma^z_4
\end{equation}
with coupling strength $J_{xx}/(2 \pi) = 5.21$ MHz, $J_{yy}/(2 \pi) = -5.20$ MHz, $J_{zz}/(2 \pi) = -9.9$ MHz and frequency shifts $\delta \omega_1/(2 \pi) = -13.41$ MHz, $\delta \omega_2/(2 \pi) = 6.77$ MHz, $\delta \omega_3/(2 \pi) = 7.38$ MHz and $\delta \omega_4/(2 \pi) = -6.96$ MHz. 

In order to estimate the effect of neglecting couplings to qubits outside the considered subsystem, we have also considered cases, where each qubit only interacts with one neighbor. This led to deviations of less than 1\% for the frequency shifts and less than 4\% for the couplings, which confirms that our test also provides a good estimate for the full lattice.

Fig.~\ref{fig:dr_qubit} shows the lowest quasienergies of the fast space of the four-body Floquet matrix and the maximal matrix element $t_{max}$ of the scattering matrix $T$ between this part of the fast subspace and the considered slow subspace. The plot shows results for two cases that result in almost the same effective couplings: the first with weaker couplings in the unmodulated lattice and stronger drive and the second in the opposite limit. 

For both cases the ratio $t_{max}/\epsilon$ never exceeds $0.1$ which confirms the validity of adiabatic elimination. This ratio is lower for the case with weak lattice coupling and strong driving, indicating that the unmodulated couplings set the limit for adiabatic elimination. Hence weakly coupled lattice with stronger driving result in a cleaner implementation.

\begin{figure}
  \centering
  \subfloat[]{\includegraphics[width=0.46\textwidth]{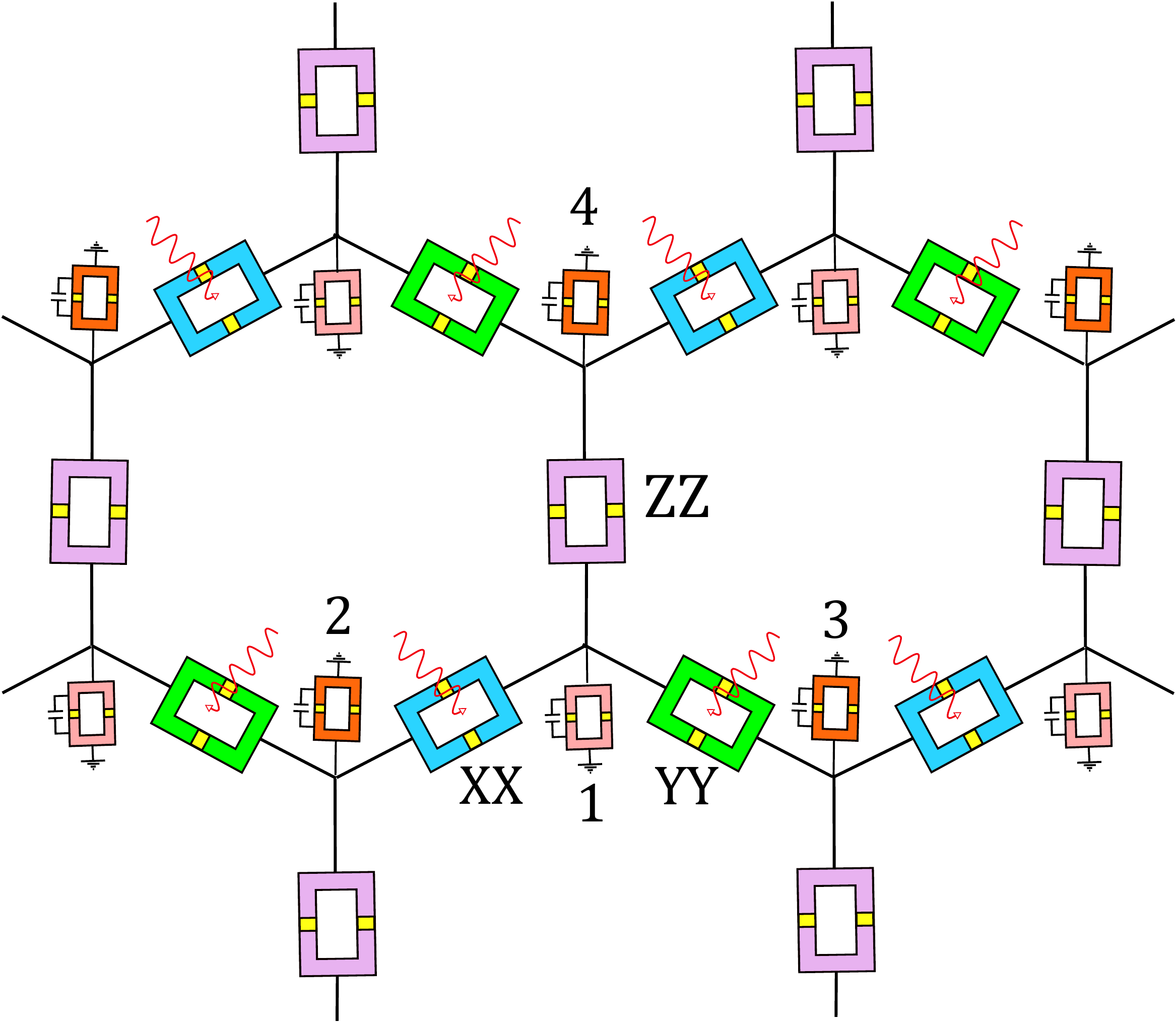}\label{fig:f1}}\\
  \subfloat[]{\includegraphics[width=0.46\textwidth]{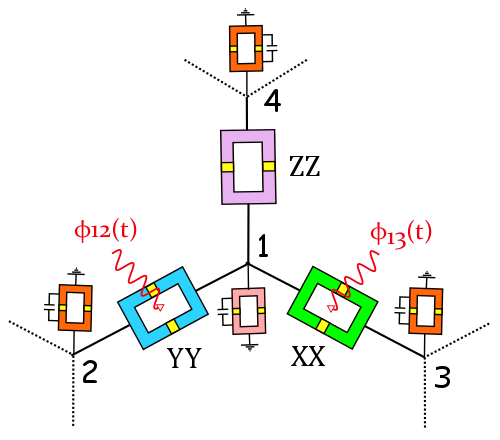}\label{fig:f2}}
  \caption{(a) Implementation of the Kitaev model on a honeycomb lattice in superconducting circuits based on driven coupling scheme (b) A typical qubit and three nearest neighbors considered for numerical simulations.}
  \label{fig:honeycomb_implementation_nonlinear}
\end{figure}

\subsection{Implementation based on driven coupluing scheme}
In the driven coupling scheme the $XX$ and $YY$ links are implemented by driven SQUIDs, c.f. Sec. \ref{sec:driven_coupling}, and the $ZZ$ links are implemented by an unmodulated SQUID, see Fig.~\ref{fig:honeycomb_implementation_nonlinear}. For the external flux through the coupler SQUID between two qubits $i$ and $j$ which form a $XX$ or $YY$ link we choose,
\begin{align}
	 \phi_{ij}(t) = \pi & + 2 \phi_{ij,1} \cos{((\omega_i - \omega_j)t + \theta_{ij})} \\
	 & + 2 \phi_{ij,2} \cos{((\omega_i + \omega_j)t)} \nn
\end{align}
where $\theta_{ij} = 0(\pi)$ for a $XX(YY)$ link.

Let us first comment on the transition frequency pattern of the qubits in this scheme. Note that as opposed to the driven qubit scheme, here the $XX$ and $YY$ interactions are first order in the coupling strength, which gives us more freedom to choose the pattern of transition frequencies. Hence, while it is possible to choose the same transition frequency pattern as for the driven qubit scheme, the transition frequencies of the qubits lying on a zigzag line formed by $XX$ and $YY$ links can also be chosen to be closer to each other. The qubits on the $ZZ$ links, in turn, are still required to be largely detuned, since the $ZZ$ interaction is fourth-order and there is a strong first-order coupling associated with this coupler. We therefore explore two regimes of qubit transition frequencies of the four-body circuit shown in Fig.~\ref{fig:honeycomb_implementation_nonlinear}.

\subsubsection{Numerical test of the approximations}
To numerically test our predictions for the driven coupling regime, we take a four-qubit circuit as shown in Fig.~\ref{fig:honeycomb_implementation_nonlinear}b and explore two patterns of transition frequencies. (1) a large detuning regime, with transition frequencies $\omega_1/(2 \pi) =  6.1$ GHz, $\omega_2/(2 \pi)=  9.6$ GHz, $\omega_3/(2 \pi)= 9.1$ GHz and $\omega_4/ (2 \pi)= 9.9$ GHz, and (2) a small detuning regime with transition frequencies $\omega_1/ (2 \pi) =  6.1$ GHz, $\omega_2/(2 \pi) =  6.45$ GHz and $\omega_3/(2 \pi)=  6.55$ GHz and $\omega_4/(2 \pi) = 9.9$ GHz. 

For both regimes we use the same coupling parameters: $g_x/(2 \pi) = 300$ MHz, $g_j/(2 \pi) = 150$ MHz, $g_z/(2 \pi) = 10$ MHz and $\phi_{ij,1} = \phi_{ij,2} = 0.1$ for the $XX$ and $YY$ couplings (here between qubits 1 and 2 and between qubits 1 and 3), and $g_x/(2 \pi) = 200$ MHz and $g_z/(2 \pi) = 10$ MHz for the $ZZ$ coupling (here between qubits 1 and 4), c.f. Eq.~(\ref{eq:bimodal_nl_Hamiltonian}). 

Performing the adiabatic elimination of the fast subspace via the Salwen method, we can verify that the effective Hamiltonian takes the same form as in Eq.~(\ref{eq:effham_4qubit}). In the large detuning regime, we get the parameters $J_{xx}/(2\pi) = J_{yy}/(2\pi) = 30$ MHz and $J_{zz}/(2\pi)=10$ MHz for the effective couplings and $\delta \omega_1/(2 \pi) = -13.92$ MHz, $\delta \omega_2/(2 \pi) = 0.22$ MHz, $\delta \omega_3/(2 \pi) = 0.24$ MHz and $\delta \omega_4/(2 \pi) = -13.88$ MHz. In the small detuning regime, we optain, $J_{xx}/(2 \pi) = 29.78$ MHz, $J_{yy}/(2 \pi) =29.80$ MHz, $J_{zz}/(2 \pi) = 9.8$ MHz, $\delta \omega_1/(2 \pi) = -14.85$ MHz, $\delta \omega_2/(2 \pi) = 0.72$ MHz, $\delta \omega_3/(2 \pi) = 0.57$ MHz and $\delta \omega_4/(2 \pi) = -14.78$ MHz.

\begin{figure*}
  \centering
  \subfloat[]{\includegraphics[width=0.46\textwidth]{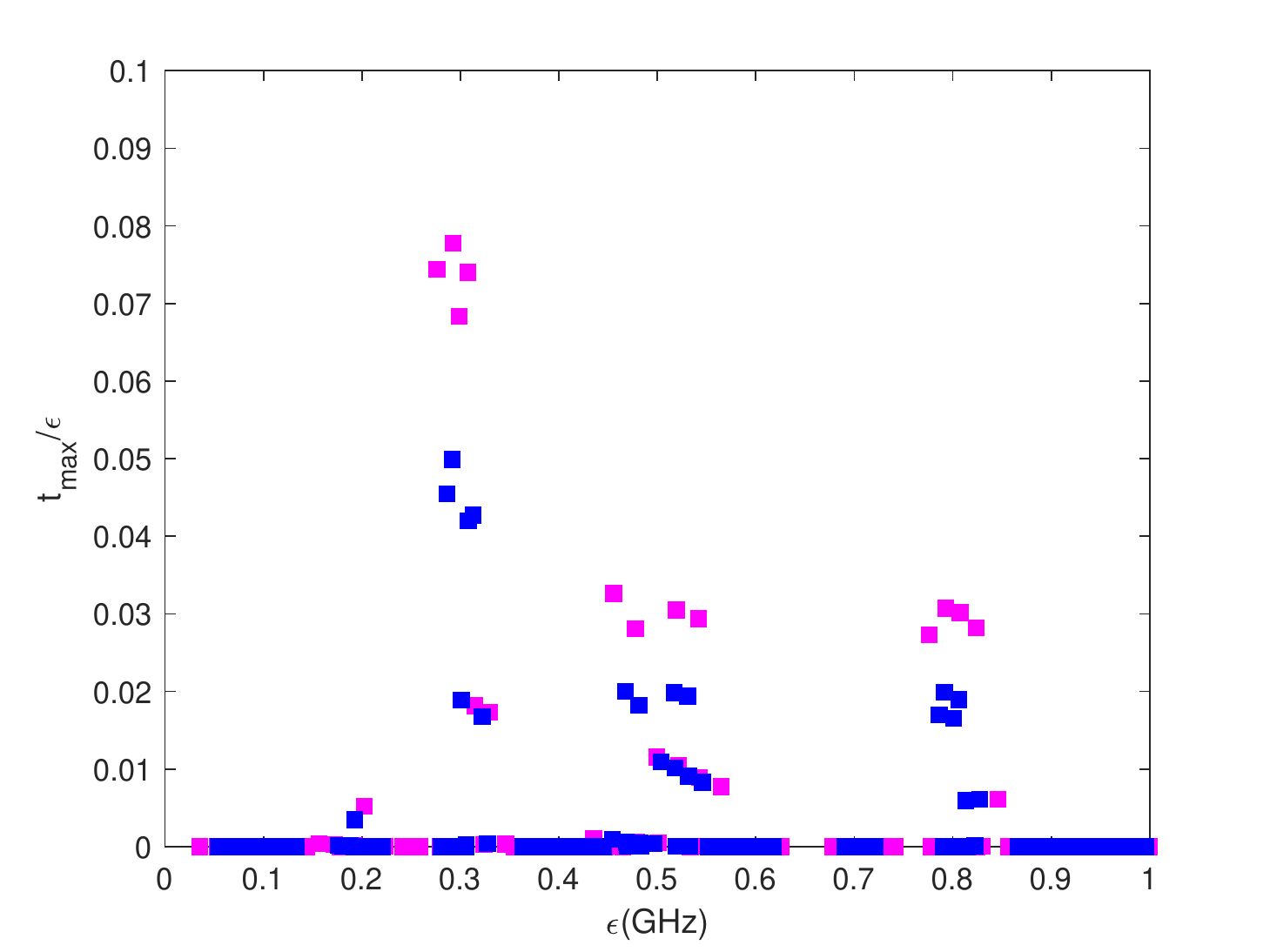}\label{fig:dr_qubit}}
  \hfill
  \subfloat[]{\includegraphics[width=0.46\textwidth]{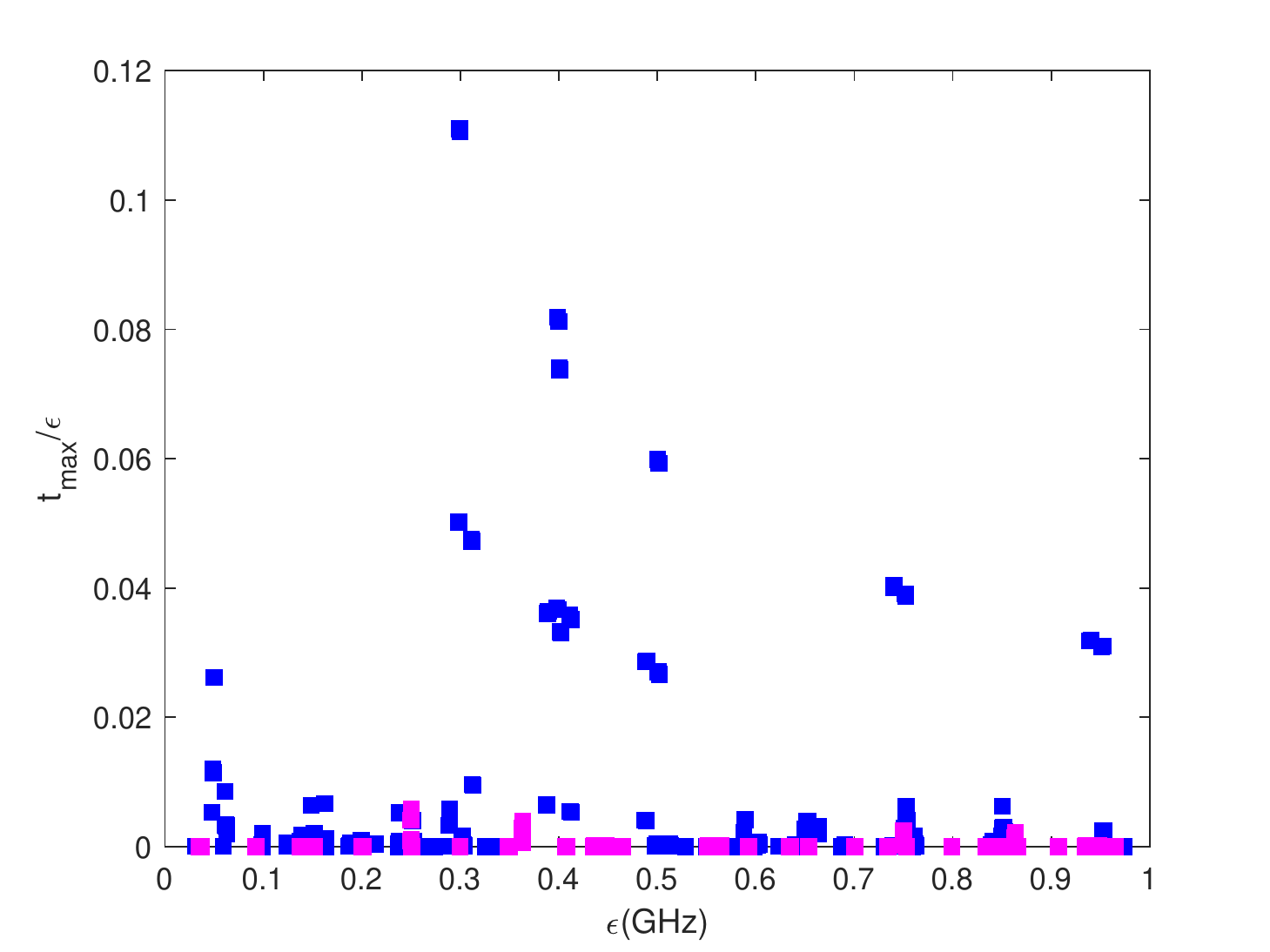}\label{fig:dr_coupling}}
 \caption{Validity of adiabatic elimination in the driven honeycomb lattice. The maximum coupling element $t_{max}$ between the slow space and the lowest states of the fast space of the driven four-body circuit is shown as a function of quasienergies of the fast space, $\epsilon$, for two driving cases. {\bf (a) Driven qubit scheme} Blue (magenta) squares show the data for weaker (stronger) couplings in the lattice with stronger (weaker) modulation. Both cases have the same effective couplings ( $\sim5.2$ MHz for $XX$ and $YY$ and $ \sim10$ MHz for $ZZ$) and an effective gap of $\sim 300$ MHz. Parameters $b_{12}/(2 \pi) = 213.3 (177.7)$ MHz, $b_{21}/(2 \pi) = -48 (-40)$ MHz, $b_{13}/(2 \pi) = -41.7 (-34)$ MHz and $b_{31}/(2 \pi) = 204.1 (166.2)$ MHz for data shown in blue (magenta). The parameters of the undriven Hamiltonian are $g_c/(2 \pi) = g_x/(2 \pi) = 200 (250)$ MHz and $g_z/(2 \pi) = 10$ MHz. {\bf (b) Driven coupling scheme} Blue (magenta) squares shows the data for small (large) detuning regime of qubit transition frequencies. Both cases have an effective coupling of $\sim 30$ MHz. For the large detuning regime the couplings are well approximated by RWA. Parameters: $\omega_1/(2 \pi)  = 6.1$ GHz, $\omega_2/(2 \pi) = 6.45$ GHz and $\omega_3/(2 \pi) =  6.55$ GHz and $\omega_4/(2 \pi)  = 9.9$ GHz for blue data and $\omega_1/(2 \pi)  =  6.1$ GHz, $\omega_2/(2 \pi)=  9.6$ GHz, $\omega_3/(2 \pi)= 9.1$ GHz and $\omega_4/(2 \pi) = 9.9$ GHz for the data in magenta.}
\end{figure*}

To validate the adiabatic elimination, we have again calculated the maximum coupling strength between the states that give rise to the effective Hamiltonian (i.e. the degenerate manifold at zero energy) and the lowest energy state in the fast space. The result is shown in Fig.~\ref{fig:dr_coupling} and shows that there is an effective gap $\sim400$MHz between the two fast and slow subspaces, while the coupling reaches a maximum of $\sim40$MHz, indicating that the effective Hamiltonian in Eq.~(\ref{eq:effham_4qubit}) indeed provides a good approximation. 

\section{Conclusions}
\label{sec:conclusions}
In conclusion, we here introduced a scheme for an analog implementation of a complete set of spin-spin interactions in circuits of coupled transmon qubits. Using the Floquet formalism for periodically and quasi-periodically modulated quantum systems, we presented a generic procedure to derive an effective Hamiltonian for the low frequency subspace of the system in the rotating frame of the qubits. We discussed several single-mode and bi-modal driving schemes to realize different spin-spin interactions in transmon circuits with driven qubits and driven nonlinear couplings.

We then combined the developed schemes for generating spin-spin interactions to show that the Kitaev honeycomb model can be generated in a lattice of coupled transmon qubits. We showed that both, driving the qubits directly and driving nonlinear couplers between them can implement the honeycomb model. In the former, the desired $XX$ and $YY$ interactions are second order processes and the strengths of these interactions were $|J_{xx}| \sim |J_{yy}| \sim 5$MHz in our numerical example. The $ZZ$ is implemented via an unmodulated SQUID and can have a strength of $|J_{zz}| \sim 10$MHz.

In the scheme with driven couplers, the $ZZ$ interaction is implemented in the same way and thus has the same strength, whereas the $XX$ and $YY$ interactions can here have a strength of $|J_{xx}| \sim |J_{yy}| \sim 30$MHz as they are first order.

It is important to note that in both schemes $J_{xx}$, $J_{yy}$ and $J_{zz}$ can be tuned independently. The first two are tuned via the modulation amplitudes and the latter is modified by the dc flux through the coupling SQUID. It is thus feasible to explore both phases of the honeycomb model, the Abelian and the non-Abelian phase \cite{Kitaev09}. 

The Abelian phase is realized when $|J_{\alpha}| > |J_\beta| + |J_\gamma|$, where $\alpha, \beta, \gamma \in \{xx,yy,zz\}$. In this phase the model can be mapped onto the toric code model \cite{Sameti16}. Using our proposed circuits, one can also investigate the non-abelian phase of the model. Compared to the abelian phase, the topological properties of this gapless phase are less explored \cite{Jiang11}. An applied magnetic field of the form, $H_B = \sum_j (h_x \sigma_j^x + h_y \sigma_j^y + h_z \sigma_j^z)$ would open a spectral gap in this phase. In our implementation scheme, this magnetic field term could be implemented by incorporating additional resonant modulations of the qubits.

Finite size effects of an implementation were predicted to become negligible for lattices of 36 spins, \cite{Kells09}.
In contrast to previously proposed implementations of the Kitaev honeycomb model \cite{You10}, our approach does not invoke an ultra-strong coupling regime, which hasn't yet been realized in multi-site lattices. Moreover, no long-range coupling terms would appear as compared to the approach in \cite{You10}.

\appendix
\section{Explicit form of Floquet matrix}
\label{sec:explicit-Floquet}
The Floquet matrix $H_F$ in Eq.~(\ref{eq:Floquet_Hamiltonian}) explicitely reads,
\begin{widetext}
\begin{equation}
H_F=
\left[
\begin{array}{c|c|c|c|c|c|c}
\ddots  & & & & & \\
\hline
\hdots & \tilde{H}^{(0)} + 2\omega_1 \mathbb{1} & \tilde{H}^{(1)} & \tilde{H}^{(2)} & \tilde{H}^{(3)} & \tilde{H}^{(4)} & \dots\\
\hline
\dots &\tilde{H}^{(-1)} & \tilde{H}^{(0)} + \omega_1 \mathbb{1} & \tilde{H}^{(1)} & \tilde{H}^{(2)} & \tilde{H}^{(3)} & \dots\\
\hline
\dots &\tilde{H}^{(-2)} & \tilde{H}^{(-1)} & \tilde{H}^{(0)} & \tilde{H}^{(1)} & \tilde{H}^{(2)} & \dots\\
\hline
\dots &\tilde{H}^{(-3)} & \tilde{H}^{(-2)} & \tilde{H}^{(-1)} & \tilde{H}^{(0)} - \omega_1 \mathbb{1} & \tilde{H}^{(1)} & \dots \\
\hline
\dots & \tilde{H}^{(-4)} &\tilde{H}^{(-3)} & \tilde{H}^{(-2)} & \tilde{H}^{(-1)} & \tilde{H}^{(0)} - 2\omega_1 \mathbb{1} & \dots \\
\hline
& & & & & & \ddots
\end{array}
\right]
\end{equation}	
\end{widetext}

\section{Projection-operator approach to Salwen perturbation theory} \label{appendix:AppB}
We here present a generalized approach to Salwen perturbation theory \cite{Salwen55}. For an unperturbed Hamiltonian $H_0$ on Hilbert space $\mathcal{H}$ with eigenstates and eigenenergies $ \ket{\psi_{\alpha}}$ and $\epsilon^{(0)}_{\alpha}$, we assume that $\mathcal{H} = S \oplus F$ in which $S$ is a the subspace of nearly degenerate eigenstates of $H_0$ such that each $\ket{\psi_{\alpha}} \in S$ is separated in energy from all other states $\ket{\psi_{\beta}} \in F$ with an energy gap, i.e.
\begin{equation}
 |\epsilon_{\alpha} - \epsilon_{\gamma}| \ll |\epsilon_{\beta} - \epsilon_{\gamma}|
\end{equation}
for any $\ket{\psi_{\gamma}} \in S$. We simply denote $\ket{\psi_{\alpha}} \in O$ by $\alpha \in O$ for $O \in S,F,\mathcal{H}$.
We now consider a perturbation $V$ to the Hamiltonian $H_0$ and write the Schr\"odinger equation for the perturbed Hamiltonian $H=H_0+V$ as,
\begin{equation} \label{eq:schr_salwen}
 H\ket{\tilde{\psi}_{\alpha}} = \epsilon_{\alpha} \ket{\tilde{\psi}_{\alpha}}, \quad \alpha \in \mathcal{H}
\end{equation}
in which $\ket{\tilde{\psi}_{\alpha}}$ is the eigenstate corresponding to $\ket{\psi_{\alpha}}$ with energy $\epsilon_{\alpha}$. The goal is to find an effective Hamiltonian $H_{eff}$ defined on $S$ which has the same eigenenergies $\epsilon_{\alpha}$ as $H$. To derive this effective description perturbatively, we assume that the energy gap is larger than any coupling element $\braket{\psi_{\alpha}|V|\psi_{\beta}}$ for $\alpha,\beta \in \mathcal{H}$. This means that the states in $S$ maintain large overlap with $S$ after turning on the interaction. We define $\mathcal{P}$ and $\mathcal{Q}$ to be the projectors onto the two subspaces $S$ and $F$ respectively,
\begin{equation}
 \mathcal{P} = \sum_{\alpha \in S} \ket{\psi_{\alpha}} \bra{\psi_{\alpha}}, \quad \mathcal{Q} = \sum_{\beta \in F} \ket{\psi_{\beta}} \bra{\psi_{\beta}},
\end{equation}
where $\mathcal{Q} = 1  - \mathcal{P}$, $\mathcal{P}^2 = \mathcal{P}$ and $\mathcal{Q}^2=\mathcal{Q}$.
By applying these projectors onto Eq.~(\ref{eq:schr_salwen}) and using $\mathcal{Q} = 1  - \mathcal{P}$, the Schr\"odinger equation can now be written as two coupled equations,
\begin{equation} \label{eq:sch_project}
 \begin{split}
  & \mathcal{P} H \mathcal{P}\ket{\tilde{\psi}_{\alpha}} + \mathcal{P} H \mathcal{Q}\ket{\tilde{\psi}_{\alpha}}= \epsilon_{\alpha} \mathcal{P} \ket{\tilde{\psi}_{\alpha}}, \\
  & \mathcal{Q} H \mathcal{P}\ket{\tilde{\psi}_{\alpha}} + \mathcal{Q} H \mathcal{Q}\ket{\tilde{\psi}_{\alpha}}= \epsilon_{\alpha} \mathcal{Q} \ket{\tilde{\psi}_{\alpha}},
 \end{split}
\end{equation}
for $\alpha \in S$.
We are just concerned with the perturbed eigenstates of $S$, so $||\mathcal{Q} \ket{\tilde{\psi}_{\alpha}} || \ll 1$. We can thus solve the second equation of (\ref{eq:sch_project}) for $\mathcal{Q} \ket{\tilde{\psi}_{\alpha}}$ to get,
\begin{equation} \label{eq:project_F}
 \mathcal{Q} \ket{\tilde{\psi}_{\alpha}} = \mathcal{G}(\epsilon_{\alpha}) \mathcal{Q} V \mathcal{P} \ket{\tilde{\psi}_{\alpha}}
\end{equation}
where we used $\mathcal{P} H_0 \mathcal{Q} = \mathcal{Q} H_0 \mathcal{P} = 0$. $\mathcal{G}(\epsilon_{\alpha}) = (\epsilon_{\alpha}  \mathbb{1} - \mathcal{Q} H \mathcal{Q})^{-1}$ is the Green's function associated with the subspace $F$ of the interacting system. 
We plug Eq.~(\ref{eq:project_F}) into Eq.~(\ref{eq:sch_project}) to get an equation which defines the effective Hamiltonian for the subspace $S$,
\begin{equation} \label{eq:eff_subspace_matrix_eq}
H_{eff}  \mathcal{P} \ket{\tilde{\psi}_{\alpha}} \equiv \epsilon_{\alpha} \mathcal{P} \ket{\tilde{\psi}_{\alpha}}
\end{equation}
with,
\begin{equation}
 H_{eff}(\epsilon_{\alpha}) \equiv \mathcal{P} H \mathcal{P} + \mathcal{P} V \mathcal{Q} \mathcal{G} (\epsilon_{\alpha})  \mathcal{Q} V \mathcal{P}
\end{equation}
One can now calculate $\mathcal{G}$ as a perturbative expansion in $V$,
\begin{equation}
 \mathcal{G}(\epsilon_{\alpha}) = \mathcal{G}_0 \sum_{n=0}^{\infty} (V \mathcal{G}_0)^n
\end{equation}
where $\mathcal{G}_0(\epsilon_{\alpha}) = \mathcal{Q}(\epsilon_{\alpha}  \mathbb{1} -  H_0)^{-1} \mathcal{Q}$ is the Green's function of the non-interacting system projected onto $F$,
\begin{equation}
 \mathcal{G}_0(\epsilon_{\alpha}) = \sum_{\beta \in F}   \frac{\ket{\psi_{\beta}} \bra{\psi_{\beta}}}{\epsilon_{\alpha} - \epsilon_{\beta}^{(0)}}
\end{equation}
The effective Hamiltonian now reads,
\begin{equation} \label{eq:eff_Hamiltonian}
 H_{eff}(\epsilon_{\alpha}) = \mathcal{P} ( H_0 + \mathcal{T}(\epsilon_{\alpha}) ) \mathcal{P}
\end{equation}
where we have defined the scattering matrix $\mathcal{T}$ as,
\begin{equation}
 \mathcal{T} (\epsilon_{\alpha}) =  V \sum_{n=0}^{\infty} (\mathcal{G}_0 V)^n
\end{equation}
 Note that the condition $||\mathcal{Q} \ket{\tilde{\psi}_{\alpha}} || \ll 1$ indicates that $|\braket{\psi_{\beta} | \mathcal{Q} | \tilde{\psi}_{\alpha} }| \approx |\braket{\psi_{\beta}|\mathcal{G}_0 V | \psi_{\alpha}}| \ll 1$, so $|\braket{\psi_{\beta}|V|\psi_{\alpha}}| \ll |\epsilon_{\beta} - \epsilon_{\alpha}|$, i.e. the coupling between the two subspaces should be much smaller than the gap as stated ealier. The effective Schr\"odinger Eq.~(\ref{eq:eff_subspace_matrix_eq}) is a self-consistent equation as the effective Hamiltonian $H_{eff}$ depends on the unknown energy $\epsilon_{\alpha}$. To get the effective Hamiltonian to second order, we can approximate $\epsilon_{\alpha} \approx \epsilon^{(0)}_{\alpha}$. For higher order corrections, one can solve for the eigenenergies of $H_{eff}(\epsilon_\alpha^{(0)})$ and plug them back to Eq.~(\ref{eq:eff_subspace_matrix_eq}) to continue recursively. We however take an alternative approach and make a perturbative ansatz for $\epsilon_{\alpha}$ as explained in the main text.

\section{Engineering interactions at multiple-photon transition} \label{ssec:multi-photon}
In Sections \ref{sssec:ax_single} and \ref{sssec:az_single}, we discussed Floquet engineering at single photon exchange which leads to effective interactions of second order. The degeneracy of low-energy states can however be achieved not only for a single photon transitions but also for multiple-photon exchange with the drive. While this typically leads to lower strength of the effective interactions, it only requires lower frequencies for the driving fields, which can be a practical advantage. We therefore briefly discuss here the engineering of interactions via multi-photon transitions.

\begin{figure*}
  \centering
  \subfloat[]{\includegraphics[height=70mm]{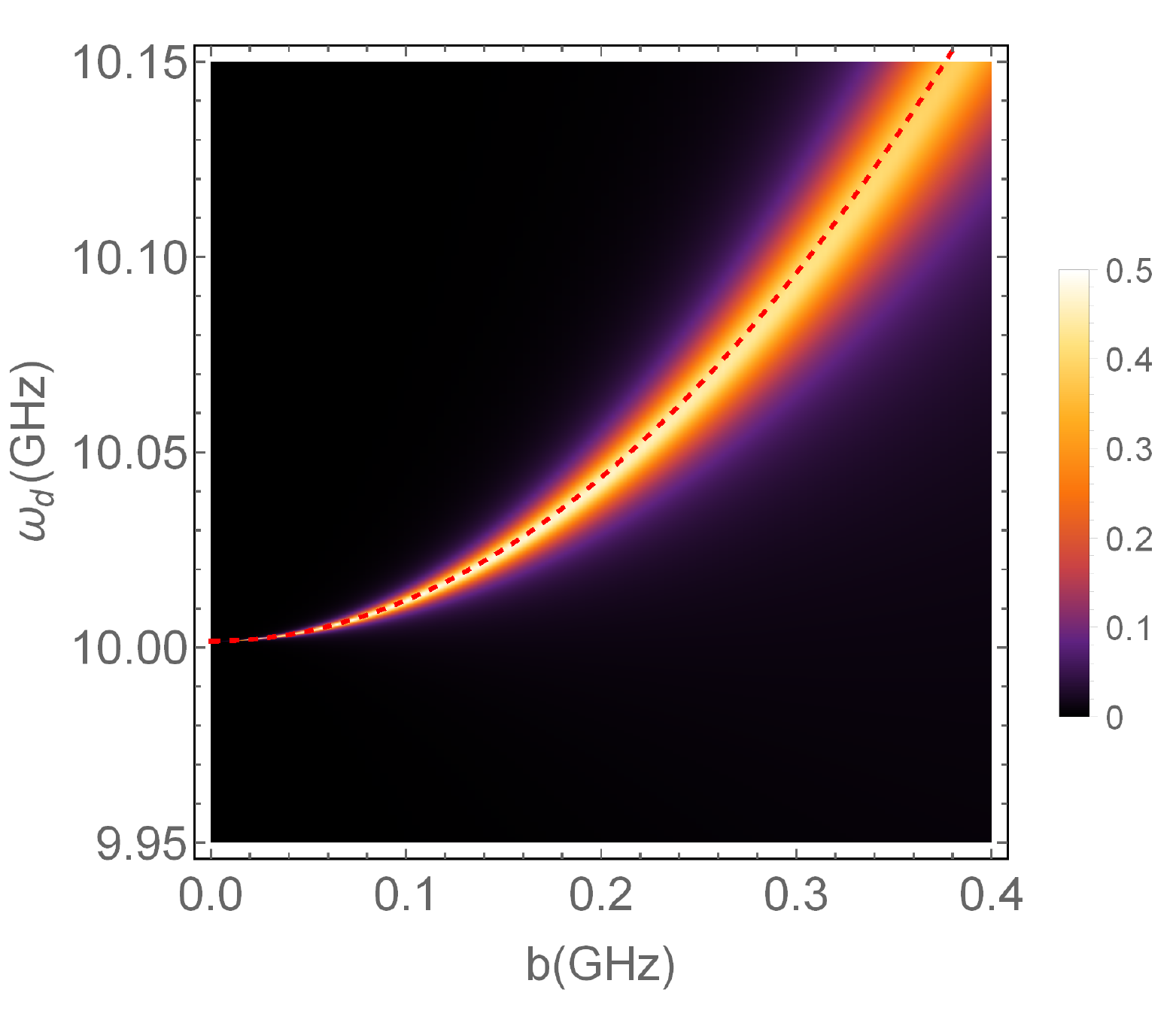}\label{fig:two_photon_map}}
  \hfill
  \subfloat[]{\includegraphics[height=70mm]{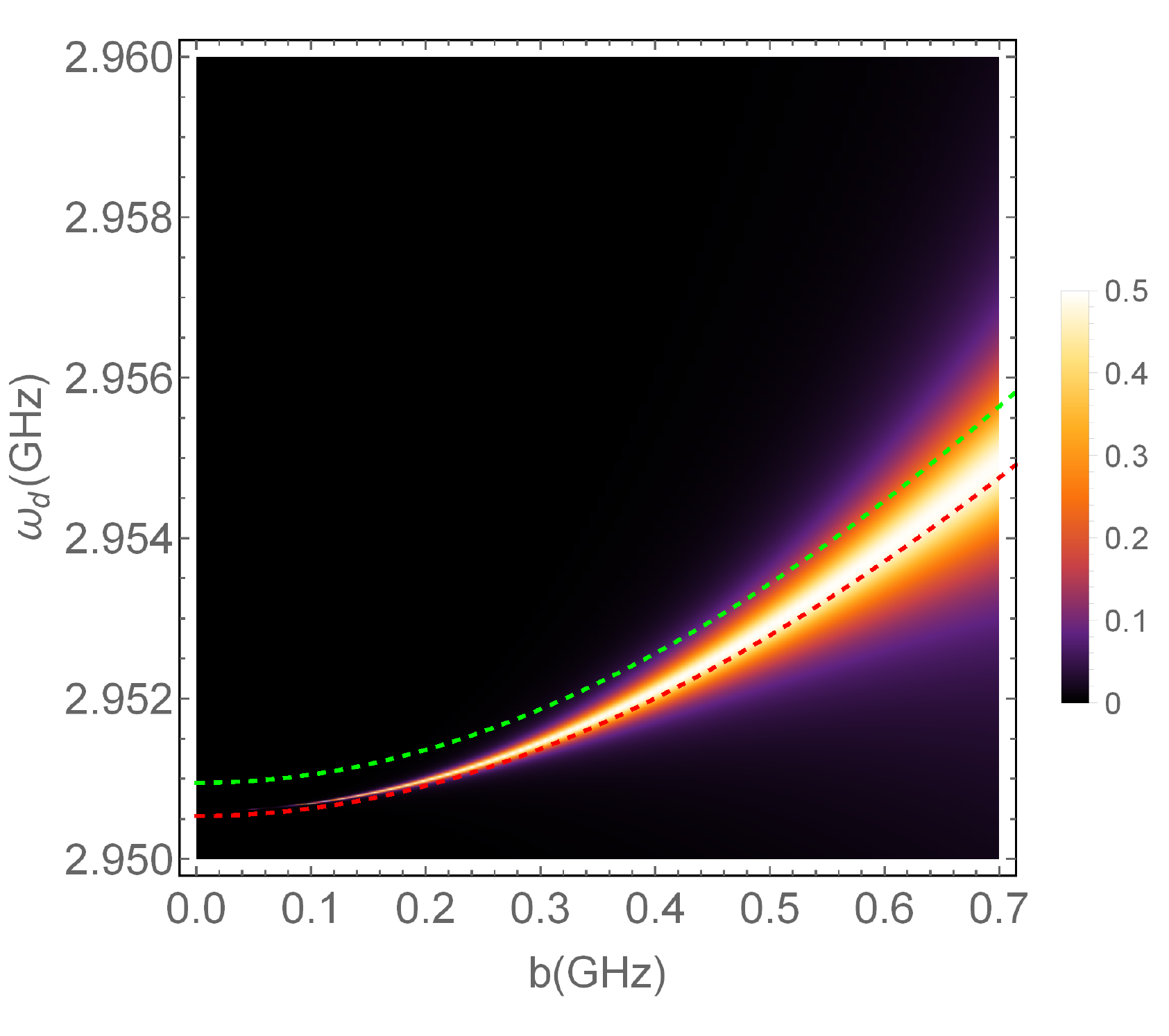}\label{fig:three_photon_map}}
  \caption{Resonances in the time-averaged transition probabilities due to spin-spin interactions with multiple photon exchange (a) Time-averaged transition probability $\ket{0 0} \rightarrow \ket{1 1}$ as a function of $\omega_d$ and $b$. The resonance is enabled by squeezing interaction at two-photon transition. The dashed line shows the fourth order analytical prediction. Parameters: $\omega_1/(2 \pi) = 11$ GHz, $\omega_2/(2 \pi)=  9$ GHz and $g_c/(2 \pi)=180$ MHz. (b) Three photon resonance due to parity-breaking interaction $\sigma_1^z \sigma_2^x$ in the time-averaged transition probability $\ket{0 0} \rightarrow \ket{0 1}$ as a function of $\omega_d$ and $b$. Dashed lines indicate the predicted resonance by the fourth (green) and sixth (red) orders of the perturbation respectively. Parameters: $\omega_1/(2 \pi) =  11$ GHz, $\omega_2/(2 \pi)= 9$ GHz and $g_c/(2 \pi)=600$  MHz.}	
\end{figure*}

\paragraph{Longitudinal modulation:}
In this case the states $\ket{\psi_1 \rangle} = \ket{\psi_1 ,  m =  -k \rangle}$, $\ket{\psi_2 \rangle} = \ket{\psi_2 , m = -k \rangle}$, $\ket{\psi_3 \rangle} = \ket{\psi_3, m =  0 \rangle}$ and $\ket{\psi_4 \rangle} = \ket{\psi_4 , m = 0 \rangle}$ are relevant for low-energy effective Hamiltonian (in the rotating frame of the qubits) when the external drive frequency is set to $\omega_d = \omega_1  \pm \omega_2/k$ ($+$ for squeezing and $-$ for hopping) with $k$ being an integer. The case of $k=1$ was discussed in Sec. \ref{sssec:az_single}. We find through the Salwen method that the effective Hamiltonian has the same structure as Eq.~(\ref{eq:structure_squeezing}) for any $k$. For $k=2$ the strengths of the squeezing and hopping read,
\begin{equation}
 J_{s} = \frac{-8 \eta b^3}{(\omega_1+\omega_2)^2} + O(b^5), \quad J_{h} = \frac{-8 \eta b^3}{(\omega_1-\omega_2)^2} + O(b^5)
\end{equation}
which are third order interactions with two-photons involved. An example of squeezing at two photon transition is shown in Fig.~\ref{fig:two_photon_map}. The parameters of the squeezing are given in table \ref{tab:par_squeezing_two_photon}. Multi-photon transitions have the advantage of using a lower drive frequency while the caveat is that the frequency $\omega_d = (\omega_1+\omega_2)/2$ is closer to the frequency of the qubits. Hence this leads to unwanted processes and in practice it is limited to lower interaction strength. For example the process $\ket{00} \rightarrow \ket{10}$ which excites just the first qubit has an average probability of $0.013$ in the given example. 

\begin{figure}[!h]
 \centering
  \subfloat[]{\includegraphics[width=0.45\textwidth]{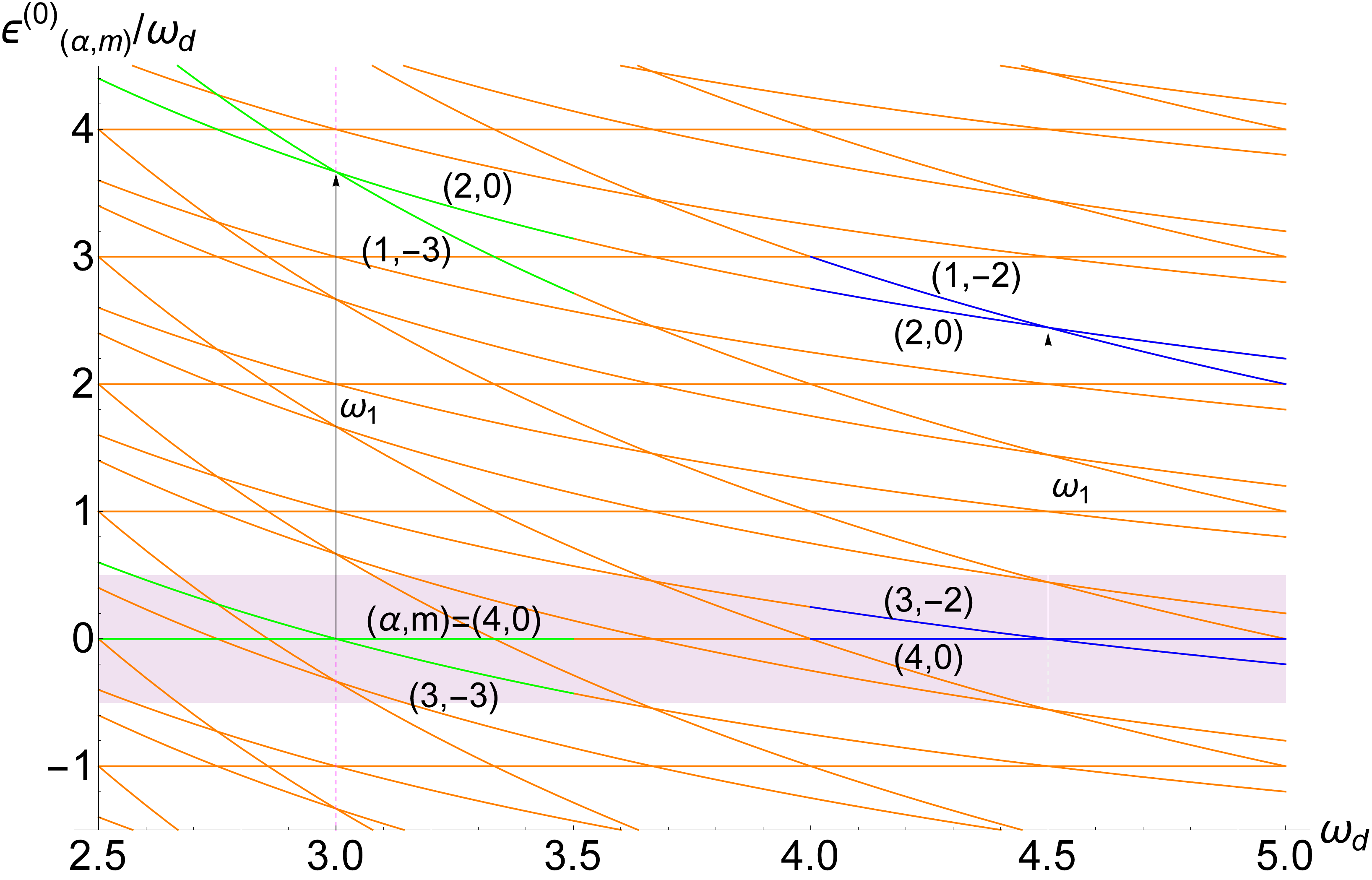}\label{fig:f1}}
  \hfill
  \subfloat[]{\includegraphics[width=0.45\textwidth]{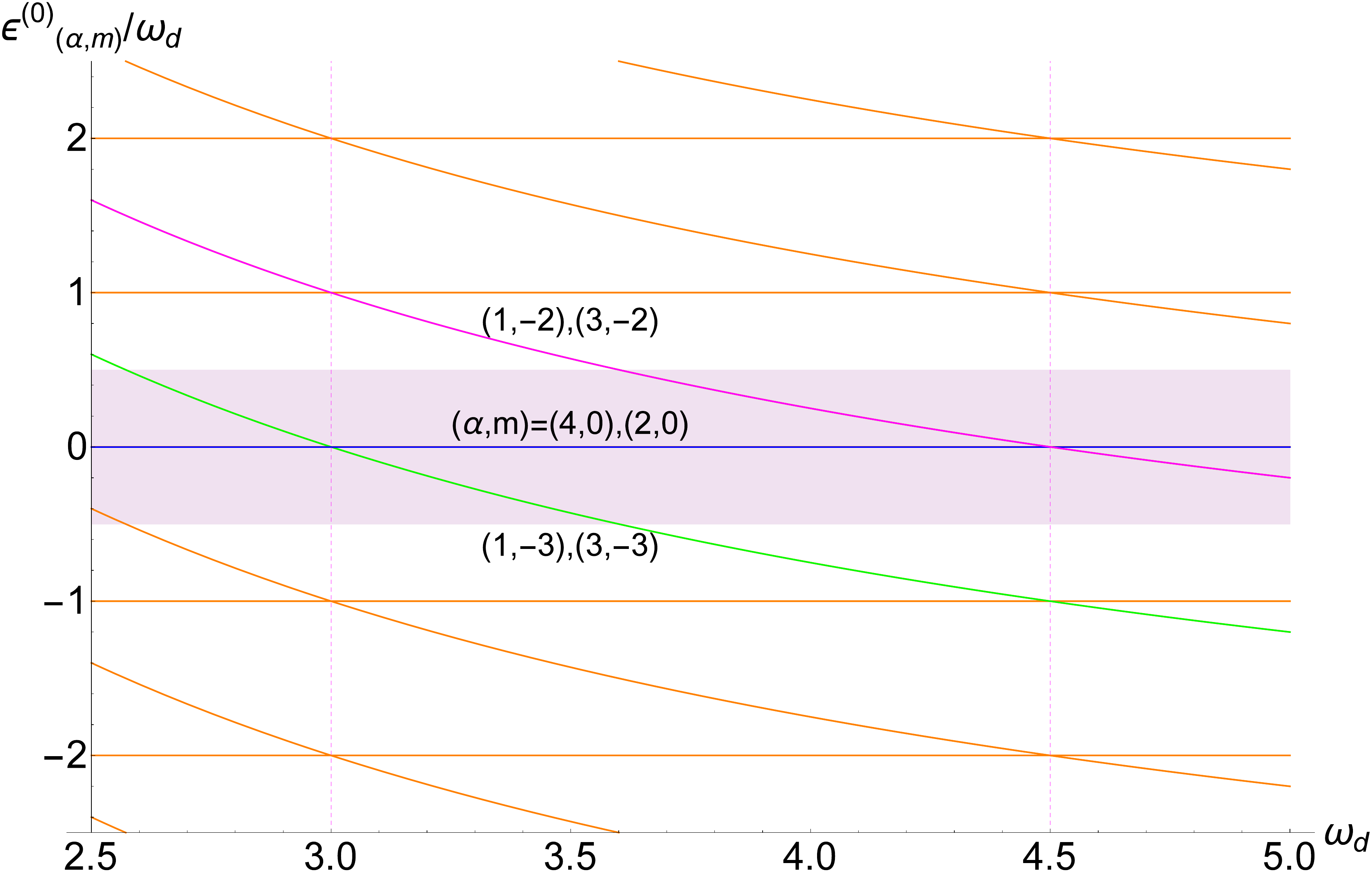}\label{fig:f2}}
  \caption{(a) Quasienergies of the Floquet matrix $H_{F0}$ as a function of $\omega_d$ for lower driving frequencies. The blue  and green curves form the degenerate manifold at $\omega_d/(2 \pi)= 4.5$ GHz and $\omega_d/(2 \pi )= 3.0$ GHz due to two and three photon transition. (b) Quasienergies in the rotating frame $U_{q1}$. The parity is restored at $\omega_d/(2 \pi )= 4.5$ GHz and the effective Hamiltonian is non-interacting. At $\omega_d/(2 \pi )= 3$ GHz, a parity-breaking term $\sigma^z_1 \sigma^x_2$ or $\sigma^z_1 \sigma^y_2$ is enabled at three-photon transition. The shaded area shows the Floquet Brillouin zone in both plots. Parameters: $\omega_1/(2 \pi) = 11$ GHz, $\omega_2/(2 \pi ) = 9$ GHz.}
  \label{fig:Floquet_eigenvalues_multiphoton}
\end{figure}

\begin{table}[tb!]
\centering
 \begin{tabular}{llll}
\hline
\multicolumn{3}{r}{Parameters (MHz)} \\
\cline{2-4}
    & $\delta \omega_1/(2 \pi)$ & $\delta \omega_2/(2 \pi)$ & $J_{s}/(2 \pi)$ \\
\hline
GSM (3rd order)      &  $23.91$     & $-7.37$   & $-2.31$     \\
GSM (4th order)       &  $23.34$     & $-7.31$   &$-2.30$       \\
GSM (6th order)       &  $23.37$     & $-7.30$   & $-2.22$      \\
Exact result       & $23.37$    & $-7.30$ & $-2.22$     \\
\hline
\end{tabular}
\caption{Numerical values for frequency shifts $\delta \omega_i$ ($i=1,2$) and squeezing interaction strength ($J_{s}$) for parameters $\omega_1/(2 \pi) = 11$ GHz, $\omega_2/(2 \pi ) = 9$ GHz, $b/(2 \pi)=120$ MHz and $\eta=1.5$ MHz. This effective interaction is constructed through two-photon exchange. GSM means Generalized Slawen method.}
\label{tab:par_squeezing_two_photon}
\end{table}
\paragraph{Transverse modulation:}
For $\omega_d = \omega_2/k$, the states $\ket{\psi_1 , m = -k \rangle}$, $\ket{\psi_2, m = 0 \rangle}$, $\ket{\psi_3 , m = -k \rangle}$ and $\ket{\psi_4 , m = 0 \rangle}$ form the degenerate manifold (see Fig.~\ref{fig:Floquet_eigenvalues_multiphoton}) and one would anticipate that the parity-breaking processes can be realized for arbitrary $k$ as in the case of $k=1$, c.f. Sec. \ref{sssec:ax_single}. However, depending on whether $k$ is odd or even the effective Hamiltonian given by the perturbation toolbox takes two different structures. For odd $k$, the Hamiltonian is block diagonal as in Eq.~(\ref{eq:structure_odd}), while for even $k$, the structure is the same as Eq.~(\ref{eq:structure_squeezing}). The reason is that the effective interaction is constructed in $(k+1)$-th order of perturbation with $k$ photons exchange. Note that applying the perturbation $H_d \sim \sigma_1^x$ an even number of times restores the parity and therefore renders effective Hamiltonian non-interacting. On the otherhand for odd $k$, one can engineer the parity breaking interactions. For $k=3$, the interaction (i.e. $J_{zx}$) is of the fourth order with the strength,
\begin{equation}
 J_{zx} = \frac{4 \eta b^4  \omega_2}{(\omega_1^2 - \omega_2^2)(\omega_1^2-\omega_2^2/9)} + O(b^6)
\end{equation}
A numerical example for $k=3$ is given in Fig.~\ref{fig:three_photon_map}. For $b=600$MHz the strength of the interacting is $1$MHz and the average occupation of the process $\ket{00} \rightarrow \ket{10}$ (closest unwanted process to resonance) is almost $0.02$.


\begin{thebibliography}{10}

\bibitem{Holthaus16}
Martin Holthaus,
{\it Floquet engineering with quasienergy bands of periodically driven optical lattices},
J. Phys. B: Atomic, Molecular and Optical Physics,
  {\bf 49}, 013001 (2016)
  
\bibitem{Oka18}
Takashi Oka and Sota Kitamura,
{\it Floquet Engineering of Quantum Materials},
arXiv:1804.03212 (2018)

\bibitem{Eckart15}
Andr\'e Eckardt and Egidijus Anisimovas,
{\it High-frequency approximation for periodically driven quantum systems from a Floquet-space perspective},
New. J. Phys. 17, 093039 (2015)

\bibitem{Goldman14}
N. Goldman and J. Dalibard,
{\it Periodically Driven Quantum Systems: Effective Hamiltonians and Engineered Gauge Fields},
Phys. Rev. X {\bf 4}, 031027 (2014)

\bibitem{Bukov15}
M. Bukov, L. D'Alessio and A. Polkovnikov,
{\it Universal high-frequency behavior of periodically driven systems: from dynamical stabilization to Floquet engineering},
Advances in Physics, Vol. {\bf 64} , Iss. 2 (2015) 

\bibitem{Goldman15}
N. Goldman, J. Dalibard, M. Aidelsburger, and N. R. Cooper
{\it Periodically driven quantum matter: The case of resonant modulations}
Phys. Rev. A 91, 033632 (2015)

\bibitem{Sameti16}
M. Sameti, A Poto\v{c}nik, D. E. Browne, A. Wallraff, and M. J. Hartmann, 
{\it Superconducting Quantum Simulator for Topological Order and the Toric Code},
Phys. Rev. A {\bf 95}, 042330 (2017). 

\bibitem{Barends14}
R. Barends,  J. Kelly,  A. Megrant,  A. Veitia,  D. Sank,  E. Jeffrey,  T. C. White,  J. Mutus,  A. G. Fowler,  B. Campbell,  Y. Chen,  Z. Chen,  B. Chiaro,  A. Dunsworth,  C. Neill,  P. O'Malley,  P. Roushan,  A. Vainsencher,  J. Wenner,  A. N. Korotkov,  A. N. Cleland, and John M. Martinis,
{\it Superconducting quantum circuits at the surface code threshold for fault tolerance}, 
Nature {\bf 508}, 500 (2014)

\bibitem{Kelly15}
J. Kelly, R. Barends, A.G. Fowler, A. Megrant, E. Jeffrey, T.C. White, D. Sank,	J.Y. Mutus,	B. Campbell, Yu Chen, Z. Chen,B. Chiaro, A. Dunsworth, I.-C. Hoi, C. Neill, P.J.J. O'Malley, C. Quintana, P. Roushan, A. Vainsencher, J. Wenner, A.N. Cleland, and John M. Martinis,
{\it State preservation by repetitive error detection in a superconducting quantum circuit}, 
Nature {\bf 519}, 66 (2015)

\bibitem{Riste14}
Rist\`e et al.,
{\it Detecting bit-flip errors in a logical qubit using stabilizer measurements},
Nat. Commun. {\bf 6}, 6983 (2015)

\bibitem{Corcoles15}
C\'{o}rcoles et al.,
{\it Detecting arbitrary quantum errors via stabilizer measurements on a sublattice of the surface code},
Nat. Comm. {\bf 6}, 6979 (2015)

\bibitem{Barends16}
Barends et al., Nature {\bf 534}, 222 (2016)

\bibitem{Ofek16}
Ofek et al.,
{\it Extending the lifetime of a quantum bit with error correction in superconducting circuits},
Nature {\bf 536}, 441 (2016)

\bibitem{Chen14}
Chen et al.,
{\it Simulating weak localization using superconducting quantum circuits}
Nat. Commun. {\bf 5}, 5184 (2014)

\bibitem{Raftery14}
J. Raftery, D. Sadri, S. Schmidt, H. E. T\"ureci, and A. A. Houck
{\it Observation of a Dissipation-Induced Classical to Quantum Transition},
Phys. Rev. X {\bf 4}, 031043 (2014)
%

\bibitem{Barends15}
R. Barends, L. Lamata, J. Kelly, L. Garc\'{i}a-\'{A}lvarez, A.G. Fowler, A. Megrant, E. Jeffrey, T. C. White, D. Sank, J. Y. Mutus, B. Campbell, Yu Chen, Z. Chen, B. Chiaro, A. Dunsworth, I.-C. Hoi, C. Neill, P. J. J. O'Malley, C. Quintana, P. Roushan, A. Vainsencher, J. Wenner, E. Solano, and John M. Martinis,
{\it Digital quantum simulation of fermionic models with a superconducting circuit},
Nat. Comm. {\bf 6}, 7654 (2015)

\bibitem{Salathe15}
Y. Salath\'e, M. Mondal, M. Oppliger, J. Heinsoo, P. Kurpiers, A. Poto\v{c}nik, A. Mezzacapo, U. Las Heras, L. Lamata, E. Solano, S. Filipp, and A. Wallraff,
{\it Digital quantum simulation of spin models with circuit quantum electrodynamics},
Phys. Rev. X {\bf 5}, 021027 (2015)

\bibitem{Eichler15}
C. Eichler, J. Mlynek, J. Butscher, P. Kurpiers, K. Hammerer, T. J. Osborne, and A. Wallraff,
{\it Exploring Interacting Quantum Many-Body Systems by Experimentally Creating Continuous Matrix Product States in Superconducting Circuits},
Phys. Rev. X {\bf 5}, 041044 (2015)

\bibitem{Roushan16}
Roushan et al., Nat. Phys. {\bf 13}, 146 (2017)

\bibitem{Houck12}
A. A. Houck, H. E. Tureci, and J. Koch,
{\it On-chip quantum simulation with superconducting circuits},
Nat. Phys. {\bf 8}, 292 (2012)

\bibitem{Fitzpatrick}
Mattias Fitzpatrick, Neereja M. Sundaresan, Andy C. Y. Li, Jens Koch, A. A. Houck
{\it Observation of a dissipative phase transition in a one-dimensional circuit QED lattice}
Phys. Rev. X {\bf 7}, 011016 (2017).

\bibitem{Roushan17}
P. Roushan et al., 
{\it Spectroscopic signatures of localization with interacting photons in superconducting qubits},
Science {\bf 358}, 1175, (2017).

\bibitem{Son09}
S.-K. Son, S. Han and S.-I. Chu,
Phys. Rev. A 79, 032301 (2009)

\bibitem{Kryiienko17}
Oleksandr Kyriienko and Anders S. S\o rensen, 
Phys. Rev. Applied 9, 064029 (2018).

\bibitem{Chunqing15}
C. Deng, J. L. Orgiazzi, F. Shen, S. Ashhab, and A. Lupascu,
Phys. Rev. Lett. 115, 133601 (2015)


\bibitem{Wendin17}
G Wendin,
Rep. Prog. Phys. 80, 106001 (2017) 


\bibitem{Shirley65}
Jon H. Shirley, 
Phys. Rev. 138, B979 (1965)

\bibitem{Ho83}
T. S. Ho, S. I Chu, J. V. Tietz,
Chem. Phys. Lett. 96, 4 (1983)

\bibitem{Leskes10}
M. Leskes, P.K. Madhu, S. Vega, 
Prog. Nucl. Magn. Reson. Spectrosc. 57(4), (2010)
\bibitem{Salwen55}
H. Salwen, Phys. Rev. {\bf 99}, 1274 (1955)

\bibitem{Floquet1883}
G. Floquet, Ann. Sci. Ecole Norm. Sup. 47, 12 (1883)

\bibitem{Sambe72}
Hideo Sambe,
Phys. Rev. A 7, 2203 (1973)

\bibitem{Kitaev03}
A. Kitaev, 
{\it Anyons in an exactly solve models and beyond},
Ann. Phys. {\bf 321}, 2 (2006).

\bibitem{Kitaev09}
A. Kitaev and C. Laumann,
{\it Topological phases and quantum computation},
arXiv:0904.2771

\bibitem{Koch07}
J. Koch, T.M. Yu, J. Gambetta, A.A. Houck, D.I. Schuster, J. Majer, A. Blais, M.H. Devoret, S.M. Girvin, and R.J. Schoelkopf, 
{\it Charge-insensitive qubit design derived from the Cooper pair box},
Phys. Rev. A {\bf 76}, 042319 (2007).

\bibitem{Kells09}
G. Kells, N. Moran and J. Vala, {\it Finite size effects in the Kitaev honeycomb lattice model on a torus}, J. Stat. Mech. P03006, (2009)

\bibitem{You10}
J. Q. You, X.-F. Shi, X. Hu, and F. Nori,
{\it Quantum emulation of a spin system with topologically protected ground states using superconducting circuits},
Phys. Rev. B {\bf 81}, 014505 (2010).

\bibitem{Jiang11}
H.-C. Jiang, Z.-C. Gu, X.-L. Qi, and S. Trebst,
{\it Possible proximity of the Mott insulating iridate Na${}_{2}$IrO${}_{3}$ to a topological phase: Phase diagram of the Heisenberg-Kitaev model in a magnetic field},
Phys. Rev. B {\bf 83}, 245104 (2011)

\bibitem{Rigetti10}
C. Rigetti and M. H. Devoret,
{\it Fully microwave-tunable universal gates in superconducting qubits with linear couplings and fixed transition frequencies}
Phys. Rev. B {\bf 81}, 134507 (2010).
 
\bibitem{Chow11}
J. M. Chow, A. D. Corcoles, J. M. Gambetta, C. Rigetti, B. R. Johnson, J. A. Smolin, J. R. Rozen, G. A. Keefe, M. B. Rothwell, M. B. Ketchen, and M. Steffen,
{\it A simple all-microwave entangling gate for fixed-frequency superconducting qubits}
Phys. Rev. Lett. {\bf 107}, 080502 (2011).

\bibitem{Reagor18}
 Matthew Reagor et al.,
 {\it Demonstration of universal parametric entangling gates on a multi-qubit lattice},
 Sci. Adv. {\bf 4}, eaao3603 (2018).
 
 \bibitem{Collodo18}
 Michele C. Collodo, Anton Poto\v{c}nik, Simone Gasparinetti, Jean-Claude Besse, Marek Pechal, Mahdi Sameti, Michael J. Hartmann, Andreas Wallraff, and Christopher Eichler,
 {\it Observation of the Crossover from Photon Ordering to Delocalization in Tunably Coupled Resonators},
 arXiv:1808.00889 (2018).


\end{thebibliography}
\end{document}